\documentclass[11pt,onecolumn,draftcls,english]{IEEEtran}

\usepackage[]{fontenc}
\usepackage[latin1]{inputenc}
\usepackage{verbatim}
\usepackage{amsmath}
\usepackage{amssymb}
\usepackage{mathrsfs}
\usepackage{graphicx}
\usepackage{amssymb}
\usepackage{color}
\usepackage{ulem}
\usepackage{xspace,bbm,epic,eepic}
\usepackage{citesort}
\usepackage[noadjust]{cite} 
\usepackage{pgfplots}
\usepackage{setspace}
\usepackage{xspace}

\title{Bayesian Pursuit Algorithms }

\author{C\' edric Herzet\textsuperscript{(1)} and Ang\'elique Dr\' emeau\textsuperscript{(2)}\\
\small \textsuperscript{(1)} INRIA-Rennes, Centre Bretagne Atlantique, Rennes, France, cedric.herzet@inira.fr \\[-0.0cm]
\small \textsuperscript{(2)} Institut Télécom, Télécom ParisTech, CNRS-LTCI, Paris, France, angelique.dremeau@telecom-paristech.fr\vspace{-0.6cm}}

\def\n{N}
\def\m{M}
\def\ss{s}
\def\xs{x}
\def\ps{p}

\def\x{{\mathbf x}}

\def\y{{\mathbf y}}
\def\s{{\mathbf s}}
\def\d{{\mathbf d}}

\def\w{{\mathbf w}}

\def\r{{\mathbf r}}

\def\hxp{\hat{\x}^{(n+1)}}

\def\D{{\mathbf D}}

\def\I{{\mathbf I}}

\def\Nc{{\cal N}}

\def\Sc{{\cal S}}

\newcommand{\Xc}{\mathcal{X}}

\def\R{\mathbb{R}}

\def\am{\arg\min}
\def\Ber{\mathrm{Ber}}
\def\vars_w{\sigma^2_n}
\def\vars{\sigma^2}
\def\ie{\textit{i.e.}, }
\def\eg{\textit{e.g.}, }
\def\etal{\textit{et al.} }

\DeclareMathOperator*{\argmin}{arg\,min}
\DeclareMathOperator*{\argmax}{arg\,max}

\begin{document}

\maketitle


\begin{abstract}
This paper addresses the sparse representation (SR) problem within a general Bayesian framework. We show that the Lagrangian formulation of the standard SR problem, \ie $\x^\star=\argmin_\x \lbrace \| \y-\D\x\|_2^2+\lambda\| \x\|_0 \rbrace$, can be regarded as a limit case of a general maximum a posteriori (MAP) problem involving Bernoulli-Gaussian variables. We then propose different  tractable implementations of this MAP problem that we refer to as ``Bayesian pursuit algorithms". The Bayesian algorithms are shown to have strong connections with several well-known pursuit algorithms of the literature (\eg MP, OMP, StOMP, CoSaMP, SP)  and generalize them in several respects. In particular,  \emph{i)} they allow for atom \emph{deselection}; \emph{ii)} they can include any prior information about the probability of occurrence of each atom within the selection process;  \emph{iii)} they can encompass the estimation of unkown model parameters into their recursions.
\end{abstract}

\section{Introduction}
Sparse representations (SR) aim at describing a signal as the combination of a small number of \emph{atoms}, namely elementary signals, chosen from a given dictionary.
More precisely, let $\y\in \R^\n$ be an observed signal  and $\D\in \R^{\n\times \m}$ a dictionary of atoms. Then, one standard  formulation of the sparse representation problem writes
\begin{align}
\x^\star &= \am_{\x}  \| \y -\D\x \|_2^2 + \lambda \| \x \|_0 , \label{eq:slsrp}
\end{align}
where $\| \cdot \|_0$ denotes the $l_0$ pseudo-norm, which counts the number of non-zero elements in $\x$, and $\lambda>0$ 
is a parameter specifying the trade-off between sparsity and distortion. 

Sparse representations have been shown to be relevant in many practical situations. A few examples include statistical regression \cite{Miller2002Subset}, digital communications \cite{Candes2005Decoding}, image processing \cite{Jeffs1993Restoration,Bobin_ieeetip07}, interpolation/extrapolation \cite{Natarajan1995Sparse}, signal deconvolution \cite{Kormylo:1982uq, Soussen2011From}, Tomo PIV \cite{Barbu2011Sparse}, compressive sampling \cite{Donoho2006Compressed}, etc.

 Unfortunately, finding the exact solution of \eqref{eq:slsrp} is a NP-hard problem \cite{Natarajan1995Sparse}, \ie it generally requires a combinatorial search over the entire solution space. 
 For problems of moderate-to-high dimensionality, combinatorial approaches are intractable and one has therefore to resort to heuristic  procedures.  
 In the current literature, three main families of algorithms can roughly be distinguished:  
 the  algorithms based on a problem relaxation,  
 the pursuit algorithms, 
 and 
 the Bayesian algorithms. 
 


The \emph{SR algorithms based on a problem relaxation} approximate the non-smooth and non-convex $\ell_0$-norm by functions easier to handle.
The resulting problem can then be solved by means of standard optimization techniques. 
 Well-known instances of algorithms based on such an approach are Basis Pursuit (BP)  \cite{Chen_siam99} and FOCUSS \cite{Gorodnitsky_ieeetsp97} which approximate the $\ell_0$-norm by the $\ell_1$-  and $\ell_p$-  ($p<1$) norms, respectively. 


The family of \emph{pursuit algorithms} encompasses all the procedures looking for a solution of the sparse representation problem by making a succession of greedy decisions on  the support \ie by iteratively selecting or deselecting  atoms from a ``local" perspective. 
%
A non-exhaustive list of algorithms belonging to this family includes matching pursuit (MP) \cite{Mallat_ieeetsp93}, orthogonal matching pursuit (OMP) \cite{Pati_asilomar93}, stagewise OMP (StOMP) \cite{Donoho_StOMP06}, orthogonal least square (OLS) \cite{Chen:1950fk}, gradient pursuit (GP) \cite{Blumensath_ieeetsp08}, iterative hard thresholding (IHT) \cite{Blumensath_jfaa08}, hard thresholding pursuit (HTP) \cite{Foucart_tr11},  compressive sampling matching pursuit (CoSaMP) \cite{Needel_ACMels08} or subspace pursuit (SP) \cite{Dai_arxiv09}. In this paper, we will more particularly focus on the family  of \emph{forward/backward} algorithms, that is procedures which consider both atom selection and deselection during the estimation process. 

Finally, \emph{Bayesian algorithms} express the SR problem as  the solution of a Bayesian estimation problem.  
 One key ingredient of the Bayesian algorithms is the choice of a proper prior, enforcing sparsity on the sought vector. 
 A popular approach consists in modelling $\x$ as a continuous random variable whose distribution has a sharp peak to zero and heavy tails \cite{Olshausen_vr97, Girolami_nc01, Fevotte_ica06, Cemgil_dsp07, Wipf:2004fj, Wipf_nips04}. 
  Another approach, recently gaining in popularity, is based on a prior made up of the combination of Bernoulli and Gaussian distributions. This model has been exploited in the following contributions \cite{Kormylo:1982uq,Larsson_ieeetsp07,Schniter_itaw08,Zayyani_icassp08,Baron_arxiv09, Herzet_icassp10, Herzet_eusipco10, Dremeau_ssp11, Qiu_arxiv11,Soussen2011From} and will be considered in this paper.

Bayesian approaches have recently gained in popularity because they allow to effectively account for uncertainties on the model parameters   or possible connections between the non-zero elements of the sparse vector (\eg in the case of structured sparsity). On the other hand, pursuit algorithms are usually attractive because of their good compromise between complexity and performance. 
   The work presented in this paper lies at the intersection of the families of Bayesian and pursuit algorithms. The contributions of the paper are threefold. First, we emphasize a connection between the standard problem \eqref{eq:slsrp} and a maximum a posteriori (MAP) problem involving Bernoulli-Gaussian (BG) variables.  In particular, we show that the set of solutions of the standard problem and the BG MAP problem are the same for certain values of the parameters of the BG model. 
Second, we propose four different procedures searching for the solution of the considered MAP estimation problem. 
Finally, we emphasize the link existing between the proposed procedures and well-known pursuit algorithms of the literature.  
 In particular, MP, OMP, StOMP and SP are shown to correspond to particular cases of the proposed algorithms for  some values of the model parameters.   

The rest of the paper is organized as follows. In section \ref{sec:connection}, we present the BG probabilistic model considered in this paper and establish a connection between \eqref{eq:slsrp} and a MAP problem involving this model. Section \ref{sec:Bpursuit} is devoted to the derivation of the proposed sparse-representation  algorithms. In section \ref{sec:biblio}, we recast our contributions within the current literature on Bayesian and pursuit algorithms; we emphasize moreover the connection between the proposed algorithms and some well-known pursuit procedures.
 Finally, in section \ref{sec:results} we provide extensive simulation results comparing, according to different figures of merit, the proposed procedures and several algorithms of the state of the art.

\section{Notations}

The notational conventions adopted in this paper are as follows. 
The $i$th element
of vector $\mathbf{a}$ is denoted $a_{i}$; 
 $\langle\mathbf{a}, \mathbf{b} \rangle\triangleq \mathbf{a}^T \mathbf{b}$ defines the scalar product between vectors $\mathbf{a}$ and  $\mathbf{b}$;   
$\|\mathbf{a}\| \triangleq \langle\mathbf{a}, \mathbf{a} \rangle^{1/2}$ is the $\ell_2$-norm of $\mathbf{a}$; $\|\mathbf{a}\|_0$ denotes the number of non-zero elements in $\mathbf{a}$. 
 The Moore-Penrose pseudo-inverse of matrix $\mathbf{A}$ is denoted by $\mathbf{A}^\dag$ and we use the notation $\I_N$ for the $N\times N$-identity matrix. 
The minimum of a function $f(\mathbf{a})$ is denoted by $\min_\mathbf{a} f(\mathbf{a})$ and  the \emph{set} of values at which this minimum is achieved by $\argmin_\mathbf{a} f(\mathbf{a})$. With a slight abuse of notation, we will often use $\mathbf{a}^\star=\argmin_\mathbf{a} f(\mathbf{a})$ to specify that $\mathbf{a}^\star$ belongs to the set of solutions, \ie $\mathbf{a}^\star\in\argmin_{\mathbf{a}} f(\mathbf{a})$.


\section{A Bayesian formulation of the standard SR problem}
\label{sec:connection}

In this section, we present the probabilistic model that will be considered throughout this paper and state a result relating the standard formulation of the SR problem \eqref{eq:slsrp} to a MAP estimation problem involving this model. 

Let $\D \in \R^{\n\times \m}$ be a dictionary whose columns are normalized to 1.  
Let moreover $\s\in\lbrace 0, 1\rbrace^M$ be a vector defining the \emph{support} of the sparse representation, \ie the subset of columns of $\D$ used to generate $\y$.  We adopt the following convention:  if $s_i=1$ (resp. $s_i=0$),  the $i$th column of $\D$ is (resp. is not) used  to form $\y$. Denoting by $\d_i$ the $i$th column of $\D$, we then consider the following observation model:    
\begin{align}
\y= \sum_{i=1}^M s_i\, x_i\,  \d_i + \w,
\end{align}   
where $\w$ is a zero-mean white Gaussian noise with variance $\vars_w$. 
Therefore,
 \begin{align}
p(\y\vert \x, \s ) &= \Nc(\D_\s \x_\s,\vars_w \mathbf{I}_N), \label{eq:obsprob}
\end{align}
where $\D_\s$ (resp. $\x_\s$) is a matrix (resp. vector) made up of the $\d_i$'s (resp. $x_i$'s) such that $s_i=1$; 
$\Nc(\boldsymbol{\mu},\boldsymbol{\Sigma})$ denotes a multivariate Gaussian distribution with mean $\boldsymbol{\mu}$ and covariance $\boldsymbol{\Sigma}$. 
 We suppose  that $\x$ and $\s$ obey the following probabilistic model:
\begin{align}
&p(\x ) = \prod_{i=1}^M p(\xs_i),
&p(\s) = \prod_{i=1}^M  p(\ss_i),
\end{align}
where
\begin{align}
p(\xs_i)&= \Nc(0,\vars_{x}),\qquad
p(\ss_i)= \Ber(p_i), \label{eq:priorsi}
\end{align}
and $\Ber(\ps_i)$ denotes a Bernoulli distribution of parameter $\ps_i$. 

 It is important to note that  \eqref{eq:obsprob}-\eqref{eq:priorsi} only define a \emph{model} on $\y$ and may not correspond to its actual distribution. Despite this fact, it is worth noticing that the BG model \eqref{eq:obsprob}-\eqref{eq:priorsi} is  well-suited to modelling situations where $\y$ stems from a sparse process. Indeed, if $\ps_i\ll 1$ $\forall\ i$, only a small number of $\ss_i$'s will \emph{typically}\footnote{In an information-theoretic sense \cite{Cover_book}, \ie  according to model \eqref{eq:obsprob}-\eqref{eq:priorsi}, a realization of  $\s$ with a few non-zero components will be observed with probability almost 1.} be non-zero, \ie the observation vector $\y$ will be generated with high probability from a small subset of the columns of $\D$.   In particular, if $p_i=p$ $\forall\ i$, typical realizations of $\y$ will involve a combination of $pM$ columns of $\D$.


We emphasize hereafter a connection between the standard problem \eqref{eq:slsrp} and a MAP problem involving model  \eqref{eq:obsprob}-\eqref{eq:priorsi}: \\[0.3cm]
\textbf{Theorem 1:}  \emph{
 Consider the following MAP estimation problem:
\begin{align}
(\hat{\x},\hat{\s})&= \argmax_{(\x,\s)} \ \log p(\y,\x,\s), \label{eq:BGp}
\end{align}
where $p(\y,\x,\s)=p(\y \vert \x, \s)\,p(\x)\, p(\s)$ is defined by the Bernoulli-Gaussian model \eqref{eq:obsprob}-\eqref{eq:priorsi}.\\
 If $\| \y \|_2< \infty$ and\vspace{-0.2cm} 
%
\begin{align}
&\vars_{x}\rightarrow \infty, \nonumber\\
&\ps_i=\ps \quad \forall\ i,  \ p\in[0,1],\label{eq:cth1}\\
&\lambda=2\vars_w \log \left(\frac{1-p}{p}\right), \nonumber
\end{align}
the BG MAP problem \eqref{eq:BGp} and the standard SR  problem \eqref{eq:slsrp} lead to the same set of solutions.}\hfill $\square$ \\[-0.1cm]

%
A proof of this result can be found in Appendix \ref{sec:Proof_Th1}.  The result established in Theorem 1 recasts the standard sparse representation problem \eqref{eq:slsrp} into a more general Bayesian framework. In particular, it shows that \eqref{eq:slsrp}  is equivalent  to a MAP estimation problem for particular values of the model parameters.  
%
In the general case, the Bayesian formulation \eqref{eq:BGp} allows for more degrees of freedom than \eqref{eq:slsrp}. For example, any prior information about the amplitude of the non-zero coefficients ($\sigma^2_x$) or the atom occurrence ($p_i$'s)  can explicitly be taken into account. 


Let us mention that a result similar to Theorem 1 was already presented in our conference paper \cite{Herzet_eusipco10} and the parallel work  by Soussen \emph{et al.} \cite{Soussen2011From}. 
 The equivalence proposed in this paper is however more general since, unlike these results,  it does not require any condition of the type
\begin{align}
\| \D_\s^\dag \y \|_0 = \|  \s\|_0 \quad\forall\, \s\in\lbrace 0, 1\rbrace^M \label{eq:condtheorem}
\end{align}
 to hold. In particular, Theorem 1 extends the equivalence between \eqref{eq:slsrp} and \eqref{eq:BGp} to the important case of noise-free data. Indeed, assume that the observed vector is generated as follows:
 \begin{align}
\y = \D_{\tilde{\s}}\tilde{\x}_{\tilde{\s}}, 
\end{align}
where $\tilde{\s}\in \{ 0, 1\}^M$,  $\|\tilde{\s}\|_0< N$, and $\tilde{\x}\in \R^M$ are realizations of some arbitrary random variables. Then,  any vector $\s$ such that 
\begin{align}
\left\{
\begin{array}{ll}
\,s_i=1& \mbox{if $\tilde{s}_i=1$},\\
\|\s\|_0 \leq N,&
\end{array}
\right. 
\end{align}
violates the equality \eqref{eq:condtheorem} and the results in  \cite{Soussen2011From,Herzet_eusipco10} do therefore not apply.

\section{Bayesian Pursuit Algorithms }
\label{sec:Bpursuit}

The BG MAP problem \eqref{eq:BGp} does not offer any advantage in terms of complexity with respect to \eqref{eq:slsrp}, since it is also NP-hard.  Hence, the resolution of \eqref{eq:BGp} requires to resort to heuristic (but practical) algorithms. In this section, we propose several greedy procedures searching for a solution of  \eqref{eq:BGp} 
 by generating a sequence of estimates $\{\hat{\x}^{(n)},\hat{\s}^{(n)}\}_{n=0}^\infty$. 
The first two procedures are particular instances of block-coordinate ascent algorithms \cite{Bertsekas1999Nonlinear}, that is, they 
 sequentially maximize the objective over subsets of elements of $\x$ and $\s$. The two other algorithms are clever heuristic procedures which do not possess any desirable ``ascent" property but rather lead to a good compromise between performance and complexity. The four procedures introduced hereafter are respectively named Bayesian Matching Pursuit (BMP), Bayesian Orthogonal Matching Pursuit (BOMP), Bayesian Stagewise Orthogonal Matching Pursuit (BStOMP) and Bayesian Subspace Pursuit (BSP) because of the clear connections existing between them and their well-known ``standard" counterparts: MP, OMP, StOMP and BSP. These connections will be emphasized and discussed in section \ref{sec:biblio}. \\[-0.2cm]

\subsubsection{Bayesian Matching Pursuit (BMP)\label{sec:BMP}}

\begin{table}[t!]
	\footnotesize{
		\fbox{
		\begin{minipage}{0.95\columnwidth}
		
		\textbf{Initialization} : $\hat{\x}^{(0)}=0$, $\hat{\s}^{(0)}=0$, $n=0$.\\
		\textbf{Repeat} : \\[0.1cm]
		$1$. Update the residual: 
		\begin{align}	
		\r^{(n)}&=\y-\D\hat{\x}^{(n)}.\label{eq:updaterestable}
		\end{align}				
		$2.$ Evaluate $\tilde{x}_i^{(n+1)}$ and $\tilde{s}_i^{(n+1)}$ $\forall i$:
		\begin{align}
		\tilde{s}_i^{(n+1)}&=
		\left\{
		\begin{array}{ll}
		1 & \mbox{if $\langle \r^{(n)}+\hat{x}_i^{(n)} \d_i, \d_i \rangle^2 > T_i $},\\
		0 & \mbox{otherwise},
		\end{array}
		\right.\label{eq:explicitBMP1table}\\
		\tilde{x}_i^{(n+1)}&= \tilde{s}_j^{(n+1)} \frac{\vars_x}{\vars_x+\vars_w}\left(\hat{x}_i^{(n)}+ \langle \r^{(n)}, \d_i \rangle\right),\label	{eq:explicitBMP2table}
		\end{align}
		with
		\begin{align}
		T_i\triangleq 2\vars_w \frac{\vars_x+\vars_w}{\vars_x} \log\left(\frac{1-p_i}{p_i}\right) .\label{eq:thresholddeftable}
		\end{align}	
		$3.$ Choose the index to be modified:\\
		\begin{align}
		j=\argmax_i \rho^{(n)}(\tilde{x}^{(n+1)}_i,\tilde{s}^{(n+1)}_i), \label{eq:index2updateminipage}
		\end{align}
		where
		\begin{align}\label{eq:residualidef}
		\rho^{(n)}(x_i,s_i)
			=& -\| \r^{(n)} +(\hat{s}_i^{(n)}\hat{x}_i^{(n)}-s_i x_i) \d_i \|^2
			 - \epsilon\, x_i^2 - \lambda_i s_i,
		\end{align}
		$4.$ Update the support and the coefficients:
		\begin{align}\label{eq:BMPSRU}
			\hat{s}_i^{(n+1)}&=
			\left\{
			\begin{array}{ll}
			  \tilde{s}_i^{(n+1)}&    \mbox{if $i=j$}, \\
			 \hat{s}_i^{(n)} &  \mbox{otherwise,} 
			\end{array}
			\right.
		\end{align}
		\begin{align}\label{eq:BMPSCU}
			\hat{x}_i^{(n+1)}&=
			\left\{
			\begin{array}{ll}
			  \tilde{x}_i^{(n+1)}&    \mbox{if $i=j$}, \\
			 \hat{x}_i^{(n)} &  \mbox{otherwise.} 
			\end{array}
			\right.
		\end{align}

		\end{minipage}
		}
	}
	\caption{\label{table:BMP} BMP algorithm}\vspace{-0.8cm}
\end{table}

We define BMP as a block-ascent algorithm in which one single component $j$ of $\hat{\x}^{(n)}$ and $\hat{\s}^{(n)}$ is modified at each iteration, that is
\begin{align}
(\hat{\x}^{(n+1)},&\hat{\s}^{(n+1)})
=\argmax_{(\x,\s)} \log p(\y, \x, \s), \label{eq:mainUpdateBMP1}
\end{align}
subject to $\forall i\neq j$:
\begin{align}
\begin{array}{l}
x_i=\hat{x}_i^{(n)},\\
s_i=\hat{s}_i^{(n)}.
\end{array}
\label{eq:mainUpdateBMP2}
\end{align}
Since only the $j$th component varies between $(\hat{\x}^{(n)},\hat{\s}^{(n)})$ and $(\hat{\x}^{(n+1)},\hat{\s}^{(n+1)})$, the update \eqref{eq:mainUpdateBMP1}-\eqref{eq:mainUpdateBMP2} is completely characterized by the value of $(\hat{x}_j^{(n+1)},\hat{s}_j^{(n+1)})$. We show in Appendix \ref{sec:derivBMP} that $\hat{x}_j^{(n+1)}$ and $\hat{s}_j^{(n+1)}$ can be expressed as
\begin{align}
\hat{s}_j^{(n+1)}&=
\left\{
\begin{array}{ll}
1 & \mbox{if $\langle \r^{(n)}+\hat{x}_j^{(n)} \d_j, \d_j \rangle^2 > T_j $},\\
0 & \mbox{otherwise},
\end{array}
\right.\label{eq:explicitBMP1}\\
\hat{x}_j^{(n+1)}&= \hat{s}_j^{(n+1)} \frac{\vars_x}{\vars_x+\vars_w}\left(\hat{x}_j^{(n)}+ \langle \r^{(n)}, \d_j \rangle\right),\label{eq:explicitBMP2}
\end{align}
where
\begin{align}
&\r^{(n)}=\y - \D \hat{\x}^{(n)},\\
&T_j = 2\vars_w \frac{\vars_x+\vars_w}{\vars_x} \log\left(\frac{1-p_j}{p_j}\right) .\label{eq:thresholddefxt}
\end{align}
A crucial question in the  implementation of \eqref{eq:mainUpdateBMP1}-\eqref{eq:mainUpdateBMP2} is the choice of the index $j$ to update at each iteration. For BMP, we choose to update the couple  $(x_j,s_j)$ leading to the maximum increase of  $\log p(\y, \x, \s)$, that is 
\begin{align}
j =\argmax_k \{ \max_{(\x,\s)} \log p(\y, \x, \s) \}, \label{eq:mainUpdateBMP3}
\end{align}
subject to \eqref{eq:mainUpdateBMP2} $\forall i\neq k$.

Equations \eqref{eq:mainUpdateBMP1}-\eqref{eq:mainUpdateBMP3} form the basis of the BMP algorithm. 
 In order to particularize these recursions to the probabilistic model defined in section \ref{sec:connection}, we define the following function:
\begin{align}
\rho^{(n)}(x_i,s_i)
=& -\| \r^{(n)} +(\hat{s}_i^{(n)}\hat{x}_i^{(n)}-s_i x_i) \d_i \|^2\nonumber\\
& - \epsilon\, x_i^2 - \lambda_i s_i, \label{eq:defrho}
\end{align}
where $\epsilon=\sigma^2_w/\sigma^2_x$ and $\lambda_i=\sigma^2_w \log((1-p_i)/p_i)$. $\rho^{(n)}(x_i,s_i)$ can be understood as the value of $ \log p(\y,\x,\s)$ (up to some additive and multiplicative constants independent of $x_i$ and $s_i$) when $(x_k,s_k)=(\hat{x}_k^{(n)},\hat{s}_k^{(n)})$ $\forall\, k\neq i$ (see \eqref{eq:defrhoanx-1}-\eqref{eq:defrhoanx0} in Appendix \ref{sec:derivBMP}). Keeping this interpretation in mind and defining
\begin{align}\label{eq:defxtildestilde}
(\tilde{x}_i^{(n+1)},\tilde{s}_i^{(n+1)})= \argmax_{(x_i,s_i)} \rho^{(n)}(x_i,s_i),
\end{align}
it is easy to see that  \eqref{eq:mainUpdateBMP3} and \eqref{eq:explicitBMP1}-\eqref{eq:explicitBMP2} can respectively be rewritten as 
\begin{align}
& j= \argmax_i \rho^{(n)}(\tilde{x}_i^{(n+1)},\tilde{s}_i^{(n+1)}),\\
&
\begin{array}{l}
\hat{s}_j^{(n+1)}=\tilde{s}_j^{(n+1)},\\
\hat{x}_j^{(n+1)}=\tilde{x}_j^{(n+1)}.
\end{array}
\end{align}
Table \ref{table:BMP} provides the analytical expressions of $\tilde{x}_i^{(n+1)}$ and $\tilde{s}_i^{(n+1)}$.  The detailed derivations leading to these expressions are provided in Appendix \ref{sec:derivBMP}. 
 A careful analysis of the operations described in Table \ref{table:BMP} reveals that BMP has a complexity per iteration scaling as $\mathcal{O}(MN)$.\\[-0.2cm]

\subsubsection{Bayesian Orthogonal Matching Pursuit (BOMP) \label{sec:BOMP}} 

\begin{table}[t!]
	\footnotesize{
		\fbox{
		\begin{minipage}{0.95\columnwidth}
		
		\textbf{Initialization} : $\hat{\x}^{(0)}=0$, $\hat{\s}^{(0)}=0$, $n=0$.\\
		\textbf{Repeat} : \\[0.1cm]
		$1$. Update the residual: 
		\begin{align}	
		\r^{(n)}=\y-\D \hat{\x}^{(n)}.\nonumber
		\end{align}	
		$2.$ Evaluate $\tilde{s}_i^{(n+1)}$ and $\tilde{x}_i^{(n+1)}$ as in \eqref{eq:explicitBMP1table}-\eqref{eq:explicitBMP2table}.\\[0.2cm]		
		$3.$ Choose $j$ as in \eqref{eq:index2updateminipage}-\eqref{eq:residualidef}.\\[0.2cm]
		$4.$ Update the support and the coefficients:
		\begin{align}\label{eq:BOMPsuppupdatetable1}
			&\quad \hat{s}_i^{(n+1)}=
			\left\{
			\begin{array}{ll}
			  \tilde{s}_i^{(n+1)}&    \mbox{if $i=j$,} \\
			 \hat{s}_i^{(n)} &  \mbox{otherwise,} 
			\end{array}
			\right.\\
		&
		\begin{array}{ll}
		\hat{\x}_{\hat{\s}^{(n+1)}}^{(n+1)}&=\left(\D^T_{\hat{\s}^{(n+1)}} \D_{\hat{\s}^{(n+1)}}+ \frac{\vars_w}{\vars_x}\I_{\| \hat{\s}^{(n+1)}\|_0} \right)^{-1}   \D^T_{\hat{\s}^{(n+1)}} \y, \\
		\hat{x}_i^{(n+1)}&=0\quad \mbox{ if $\hat{s}_i^{(n+1)}=0$.} \label{eq:BOMPcoefupdatetable}
		\end{array}
		\end{align}
		\end{minipage}
		}
	}
	\caption{\label{Table:BOMP} BOMP algorithm}\vspace{-0.8cm}
\end{table}

as BMP, we define BOMP as a particular instance of a block-coordinate ascent algorithm applied to \eqref{eq:BGp}. The subsets of variables with respect to which the objective is sequentially optimized  differ however from those considered by BMP.  In particular, we define the BOMP recursions (by means of half iterations) as follows:
\begin{align}
(\hat{\x}^{(n+\frac{1}{2})}, \hat{\s}^{(n+\frac{1}{2})})
=\argmax_{(\x,\s)} \log p(\y, \x, \s), \label{eq:mainUpdateBOMP1}
\end{align}
subject to \eqref{eq:mainUpdateBMP2} $\forall i\neq j$, with $j$ defined in \eqref{eq:mainUpdateBMP3}, then 
\begin{align}
\hat{\s}^{(n+1)}&=\hat{\s}^{(n+\frac{1}{2})},\\
\hat{\x}^{(n+1)}
&=\argmax_{\x} \log p(\y, \x, \hat{\s}^{(n+1)}) \label{eq:mainUpdateBOMP2}.
\end{align}
BOMP is therefore a two-step procedure: in a first step BOMP optimizes the goal function with respect to a particular couple $(x_j,s_j)$; this operation is strictly equivalent to BMP's recursion \eqref{eq:mainUpdateBMP1}-\eqref{eq:mainUpdateBMP2}. In a second step, BOMP looks for the maximum of $\log p(\y,\x,\s)$ over $\x$ while $\s=\hat{\s}^{(n+1)}=\hat{\s}^{(n+\frac{1}{2})}$. The solution of this problem can be expressed as
\begin{align}\nonumber
\begin{array}{l}
\hat{\x}_{\hat{\s}^{(n+1)}}^{(n+1)}=\left(\D^T_{\hat{\s}^{(n+1)}} \D_{\hat{\s}^{(n+1)}}+ \frac{\vars_w}{\vars_x}\I_{\| \hat{\s}^{(n+1)}\|_0} \right)^{-1}   \D^T_{\hat{\s}^{(n+1)}} \y, \\
\hat{x}_i^{(n+1)}=0\quad \mbox{ if $\hat{s}_i^{(n+1)}=0$.}
\end{array}
\end{align}
We refer the reader to Appendix \ref{sec:derivBMP} for a detailed derivation of this expression. 

 Particularizing \eqref{eq:mainUpdateBOMP1}-\eqref{eq:mainUpdateBOMP2} to the probabilistic model presented in section \ref{sec:connection}, we obtain the implementation described in Table  \ref{Table:BOMP}. 
 In our description, we used the fact that step \eqref{eq:mainUpdateBOMP1} is strictly equivalent to \eqref{eq:mainUpdateBMP1}-\eqref{eq:mainUpdateBMP2} and can therefore be efficiently implemented as described in the previous section. Moreover, the value of 
 $\hat{\x}^{(n+\frac{1}{2})}$ does not need to be explicitly evaluated since it is never used for the evaluation of $\hat{\x}^{(n+1)}$ and $\hat{\s}^{(n+1)}$. 
 The crucial difference between BMP and BOMP lies in the coefficient update:  \eqref{eq:BMPSCU} in BMP is replaced by \eqref{eq:BOMPcoefupdatetable} in BOMP; this alternative update has a larger computational load ($\mathcal{O}(\| \hat{\s}^{(n+1)}\|_0^3)$  for \eqref{eq:BOMPcoefupdatetable} against $\mathcal{O}(1)$ for \eqref{eq:BMPSCU}). The complexity per iteration of BOMP is therefore $\mathcal{O}(\| \hat{\s}^{(n+1)}\|_0^3+MN)$. \\[-0.2cm]
 

\subsubsection{Bayesian Stagewise Orthogonal Matching Pursuit (BStOMP) \label{sec:BStOMP}} 

We define BStOMP as a modified version of BOMP where several entries of the support vector $\s$ can be changed at each iteration. 
 In particular, BStOMP is characterized by the following recursion:
\begin{align} 
\hat{s}_i^{(n+1)}&= \tilde{s}_i^{(n+1)}\quad \forall\ i,\label{eq:thresholdStOMP0}\\
\hat{\x}^{(n+1)}&=\argmax_{\x} \log p(\y, \x, \hat{\s}^{(n+1)}).
\end{align}
where $\tilde{s}_i^{(n+1)}$ has been defined in \eqref{eq:defxtildestilde}. We remind the reader that $\tilde{s}_i^{(n+1)}$ corresponds to the optimal decision on $s_i$ when $\log p(\y, \x, \s)$ is optimized over $(x_i, s_i)$ while $(x_k, s_k)=(\hat{x}_k^{(n)}, \hat{s}_k^{(n)})$ $\forall k\neq i$. $\tilde{s}_i^{(n+1)}$ can therefore be understood as the locally-optimal decision on $s_i$ given the current estimate $(\hat{\x}^{(n)},\hat{\s}^{(n)})$. Hence, in a nutshell, the philosophy behind BStOMP consists in setting each element of $\hat{\s}^{(n+1)}$ to its locally-optimal value given the current estimate $(\hat{\x}^{(n)},\hat{\s}^{(n)})$. The update of $\hat{\x}^{(n+1)}$ is the same as for BOMP.

The operations performed by BStOMP are summarized in Table \ref{Table:BStOMP}. The complexity per iteration of BStOMP is similar to BOMP, that is $\mathcal{O}(\| \hat{\s}^{(n+1)}\|_0^3+MN)$.  In fact,  BOMP and BStOMP only differ in the support update step: whereas BOMP only sets one element of $\hat{\s}^{(n)}$ to its locally-optimal value (see \eqref{eq:BOMPsuppupdatetable1}), BStOMP does so for all components of the new support estimate (see \eqref{eq:BStOMPcoefupdatetable1}).  A consequence of update \eqref{eq:BStOMPcoefupdatetable1} is that BStOMP is no longer an ascent algorithm. 
 Nevertheless, we will see in the empirical results presented in section \ref{sec:results} that the support update implemented by BStOMP allows for a  reduction of the number of iterations (and hence of the overall running time) required to find the sought sparse vector. \\[-0.2cm]
\begin{table}[t!]
	\footnotesize{
		\fbox{
		\begin{minipage}{0.95\columnwidth}
		
		\textbf{Initialization} : $\hat{\x}^{(0)}=0$, $\hat{\s}^{(0)}=0$, $n=0$.\\
		\textbf{Repeat} : \\[0.1cm]
		$1$. Update the residual: 
		\begin{align}	
		\r^{(n)}=\y-\D \hat{\x}^{(n)}.\nonumber
		\end{align}	
		$2.$ Evaluate $\tilde{s}_i^{(n+1)}$ as in \eqref{eq:explicitBMP1table}\\[0.2cm]		
		$3.$ Update the support and the coefficients:
		\begin{align}\label{eq:BStOMPcoefupdatetable1}
			&\quad \hat{s}_i^{(n+1)}=
			  \tilde{s}_i^{(n+1)}\quad \forall\,i,\\
		&
		\begin{array}{ll}
		\hat{\x}_{\hat{\s}^{(n+1)}}^{(n+1)}&=\left(\D^T_{\hat{\s}^{(n+1)}} \D_{\hat{\s}^{(n+1)}}+ \frac{\vars_w}{\vars_x}\I_{\| \hat{\s}^{(n+1)}\|_0} \right)^{-1}   \D^T_{\hat{\s}^{(n+1)}} \y,\\
		\hat{x}_i^{(n+1)}&=0\quad \mbox{ if $\hat{s}_i^{(n+1)}=0$.} \label{eq:BStOMPcoefupdatetable2}
		\end{array}
		\end{align}
		\end{minipage}
		}
	}
	\caption{\label{Table:BStOMP} BStOMP algorithm}\vspace{-0.8cm}
\end{table}

\subsubsection{Bayesian Subspace Pursuit (BSP) \label{BSP}}

We define BSP as another heuristic procedure in which a limited number of atoms can be selected or deselected at each iteration. As previously, the choice of the atoms selected/deselected is made from a ``local" perspective. More formally, let us define
\begin{align}\label{eq:defrhosi}
\rho_i^{(n)}(s_i)&\triangleq \max_{x_i}\rho^{(n)}(x_i,s_i).
\end{align}
Hence, $\rho_i^{(n)}(s_i)$ corresponds to the maximum of $\log p(\y,\x,\s)$ when optimized over $x_i$ for a given value of $s_i$ and for $(x_k,s_k)=(\hat{x}_k,\hat{s}_k)$ $\forall k\neq i$. 
 Using this definition, we define BSP as the following  two-step procedure. First, the support estimate is updated as
\begin{align} 
\hat{\s}^{(n+\frac{1}{2})}&= \argmax_{\s\in\Sc_P} \sum_i \rho_i^{(n)}(s_i),\label{eq:BMPsu1}
\end{align}
where $\Sc_P =\{ \s\, \vert \, \| \s-\hat{\s}^{(n)}\|_0 \leq P\}$,  
and $\hat{\x}^{(n+\frac{1}{2})}$ is computed as in \eqref{eq:BOMPcoefupdatetable}. Then, in a second step, the support estimate is modified according to
\begin{align} 
\hat{\s}^{(n+1)}&= \argmax_{\s\in\Sc_K} \sum_i \rho_i^{(n+\frac{1}{2})}(s_i),\label{eq:BMPsu2}
\end{align}
where $\Sc_K =\{ \s\, \vert \, \| \s\|_0 = K\}$ and
the coefficient estimate $\hat{\x}^{(n+1)}$ is again computed from \eqref{eq:BOMPcoefupdatetable}. 

 In a nutshell, update \eqref{eq:BMPsu1} consists in selecting/deselecting the (at most) $P$ atoms leading to the best local increases of $\log p(\y, \x,\s)$ around $(\hat{\x}^{(n)},\hat{\s}^{(n)})$. This operation can be interpreted as an intermediate between BOMP and BStOMP support updates. In particular,  if $P=1$ (resp. $P=M$)  one recovers BMP/BOMP (resp. BStOMP) update \eqref{eq:BMPSRU}
 (resp. \eqref{eq:thresholdStOMP0}). In a second step, BSP modifies the support on the basis of the local variations of $\log p(\y, \x,\s)$ around $(\hat{\x}^{(n+\frac{1}{2})},\hat{\s}^{(n+\frac{1}{2})})$ with the constraint that the new support has exactly $K$ non-zero elements. 

We show in Appendix \ref{sec:derivBMP} that $\rho_i^{(n)}(s_i)= \rho^{(n)}(\tilde{x}_i(s_i),s_i)$ where 
\begin{align}
\tilde{x}_i(s_i) = s_i \, \frac{\vars_x}{\vars_x+\vars_w}\left(\hat{x}_i^{(n)}+ \langle \r^{(n)}, \d_i \rangle\right).
\end{align}
Remember that $\rho^{(n)}(x_i,s_i)$  $\forall\, i$ can be evaluated with a complexity $\mathcal{O}(MN)$. 
Moreover, we emphasize in Appendix \ref{sec:derivBMP} that solving \eqref{eq:BMPsu1}-\eqref{eq:BMPsu2} essentially requires the sorting of $L (\leq M)$ metrics depending on $\rho^{(n)}(s_i)$. The complexity associated to this operation scales as $\mathcal{O}(L \log L)$. 
Hence, the complexity per iteration of BSP is similar to BOMP and BStOMP.

\subsubsection{Parameter estimation and adaptive threshold \label{sec:BayEfroymson}}

 we now discuss the implementation of the estimation of the noise variance $\sigma^2_w$ into the iterative process defined by the Bayesian pursuit algorithms. 
At each iteration, we can consider the following maximum-likelihood estimate
\begin{align}\label{eq:MLnoisevarest}
\hat{(\vars_w)}^{(n)}
&= \arg\max_{\vars_w} \log p(\y,\hat{\x}^{(n)},\hat{\s}^{(n)}),\\
&= N^{-1} \| \r^{(n-1)} \|^2.\label{eq:varestimate}
\end{align}
This estimate can be included within the pursuit recursions defined in the previous subsections. From a practical point of view, the algorithms described in Table \ref{table:BMP} to \ref{Table:BStOMP} then remain identical but, at each iteration,  $\vars_w$ is replaced by its current estimate $\hat{(\vars_w)}^{(n)}$. In particular, BMP and BOMP remains ascent algorithms since \eqref{eq:MLnoisevarest} defines an ascent operation. 


It is illuminating to focus in more details on the impact of the estimation of the noise variance on the  update of the support $\hat{\s}^{(n)}$. 
In particular, replacing  $\vars_w$ by its estimate \eqref{eq:varestimate} leads to the following expression for $T_i$:
\begin{align}
T_i^{(n)}\triangleq& 2 \frac{\| \r^{(n)} \|^2}{N} \,  \log\left(\frac{1-p_i}{p_i}\right)
\frac{\vars_x+ N^{-1} \| \r^{(n)} \|^2 }{\vars_x} .\label{eq:thresholdadapt}
\end{align}
The threshold therefore becomes a function of the number of iterations. Moreover, as $\vars_x\rightarrow \infty$, \eqref{eq:thresholdadapt} tends to:
\begin{align}
T_i^{(n)}\stackrel{\vars_x\rightarrow \infty}{=} 2  \frac{\| \r^{(n)}\|^2}{N} \log\left(\frac{1-p_i}{p_i}\right) . \label{eq:thresholdadaptvraxinf}
\end{align}
The threshold is then proportional to the residual energy; the proportionality factor depends on the occurrence probability of each atom.
In practice, $T_i^{(n)}$ has therefore the following operational meaning: during the first iterations, the residual is large (and so is $T_i^{(n)}$), and only the atoms having a large correlation with $\y$ are likely to be included in the support; after a few iterations, the norm of the residual error decreases and atoms weighted by smaller coefficients can enter the support.


\section{Connections with previous works}\label{sec:biblio}

The derivation of practical and effective algorithms searching for a solution of the sparse problem has been an active field of research for several decades. 
 In order to properly replace our work in the ever-growing literature pertaining to this topic, we provide hereafter a short survey of some significant works in the domain. Note that, although our survey will necessarily be incomplete, we attempted to present the works the most connected with the proposed methodologies.   In the first two subsections, we review the procedures belonging to the family of pursuit and Bayesian algorithms, respectively. In the last subsection, we emphasize some nice connections existing between the proposed procedures and some well-known pursuit algorithms of the literature, namely MP, OMP, StOMP and SP.

\subsection{Pursuit algorithms}
The designation ``pursuit algorithms" generally refers to procedures looking for a sparse vector minimizing a goal function (most often the residual error $\r^{(n)}$) by making a succession of locally-optimal decisions on the support. The family of pursuit algorithms has a long history which traces back to 60's, for instance in the field of statistical regression \cite{Miller_book}. 

Within this family, one can distinguish between \emph{forward}, \emph{backward} and \emph{forward/backward} procedures.  
\emph{Forward} algorithms gradually increase the support by sequentially \emph{adding} new atoms.  In this family, one can mention matching pursuit (MP) \cite{Mallat_ieeetsp93}, orthogonal matching pursuit (OMP) \cite{Pati_asilomar93}, stagewise OMP (StOMP) \cite{Donoho_StOMP06}, orthogonal least square (OLS) \cite{Chen:1950fk} or gradient pursuit (GP) \cite{Blumensath_ieeetsp08}. These algorithms essentially differ in the way they select the atoms to be included in the support and/or  the way they update the value of the non-zero coefficients. 

\emph{Backward} algorithms use the opposite strategy: they start from a support containing all the atoms of the dictionary and reduce it by sequentially removing "irrelevant" atoms. 
 Backward algorithms have been extensively studied for undercomplete dictionaries  in the statistical regression community \cite{Miller_book}. They have been revisited more recently by Couvreur \etal in \cite{Couvreur:2000:OBG:347291.347308}. They are however of poor interest in overcomplete settings since most of them cannot make any relevant decision as soon as $N<M$.

Finally, \emph{forward/backward} algorithms make iteratively a new decision on the support of the sparse vector by \emph{either} adding and/or removing atoms from the current support. The first \emph{forward/backward} algorithm we are aware of is due to Efroymson \cite{Efroymson_60} and was placed in the context of statistical regression in undercomplete dictionaries. 
 In his paper, the author suggested to add (resp. remove) \emph{one} atom from the support if the decision leads to a residual error above (resp. below) a prespecified threshold. The choice of the threshold derives from considerations based on statistical hypothesis testing. Variations on this idea has been proposed in \cite{Berk1980Forward, Broersen_jrssc86} where the authors suggest different testing approaches. 
 
 Efroymson's procedure has later on been revisited in the context of sparse representations in overcomplete dictionary, see \eg \cite{Haugland_springer07, Zhang_ieeeit11}. Other procedures, more flexible in the number of atoms added or removed from the support have been recently published. Let us mention the iterative hard thresholding (IHT) \cite{Blumensath_jfaa08}, hard thresholding pursuit (HTP) \cite{Foucart_tr11}, compressive sampling matching pursuit (CoSaMP) \cite{Needel_ACMels08} and subspace pursuit (SP) \cite{Dai_arxiv09}. 

The procedures derived in section \ref{sec:Bpursuit} can be cast within the family of forward-backward pursuit algorithms since they build their estimate by a sequence of locally-optimal decisions and allow for both atom selection and deselection. However, unlike the proposed algorithms, most of the forward-backward procedures (\eg \cite{Efroymson_60,Berk1980Forward,Blumensath_jfaa08,Foucart_tr11,Dai_arxiv09}) do not derive from an optimization problem but are rather clever heuristic methodologies. 
Moreover, the Bayesian framework in which the proposed methods arise can account for different prior information on the  atom occurrence and encompass the estimation of some unknown model parameters. This is in contrast with the deterministic settings from which standard forward/backward algorithms derive. 

\subsection{Bayesian algorithms}

Apart from some noticeable exceptions (\eg \cite{Kormylo:1982uq}), the development of sparse representation algorithms based on Bayesian methodologies seems to be more recent. 
 The Bayesian algorithms available in the literature mainly differ in three respects: \emph{i)} the probabilistic model they use to enforce sparsity; \emph{ii)} the Bayesian criterion they intend to optimize (\eg minimum mean square error (MMSE), maximum a posteriori (MAP), etc.); \emph{iii)}  the practical procedure they apply to compute or approximate the sought solution (\eg gradient algorithm, variational approximation, Markov-Chain Monte-Carlo methods, etc.).
  Regarding the choice of the prior, a popular approach consists in modelling $\x$ as a continuous random variable whose distribution has a sharp peak to zero and heavy tails (\eg Laplace, $t$-Student or Jeyffrey's distributions). Such a strategy has been exploited, considering different Bayesian criteria and optimization strategies, in the following contributions \cite{Olshausen_vr97, Girolami_nc01, Fevotte_ica06, Cemgil_dsp07, Wipf:2004fj, Wipf_nips04}. 
  Another approach, recently gaining in popularity, is based on a prior made up of the combination of Bernoulli and Gaussian distributions, see \eg \cite{Kormylo:1982uq,Larsson_ieeetsp07,Schniter_itaw08,Zayyani_icassp08,Baron_arxiv09, Herzet_icassp10, Herzet_eusipco10, Dremeau_ssp11, Qiu_arxiv11,Soussen2011From}.
   Different variants of Bernoulli-Gaussian (BG) models exist. A first approach consists in modelling the elements of $\x$ as Gaussian variables   whose variance is controlled by a Bernoulli variable: 
  \begin{align}
p(\y\vert \x) &= \Nc(\D \x,\vars_w \mathbf{I}_N), \label{eq:obsprob2}
\end{align}
\begin{align}
&p(\x\vert \s ) = \prod_{i=1}^M p(\xs_i \vert \ss_i ),
&p(\s) = \prod_{i=1}^M  p(\ss_i),
\end{align}
where
\begin{align}
p(\xs_i\vert \ss_i)&= \Nc(0,\vars_{x}(\ss_i)),\qquad
p(\ss_i)= \Ber(p_i). \label{eq:priorsi2}
\end{align}
Hence a small variance $\vars_{x}(0)$ enforces $x_i$ to be close to zero if $s_i=0$. Another model based on BG variables is \eqref{eq:obsprob}-\eqref{eq:priorsi}, as considered in the present paper. 
Note that although models \eqref{eq:obsprob}-\eqref{eq:priorsi} and \eqref{eq:obsprob2}-\eqref{eq:priorsi2} are usually both referred to as ``Bernoulli-Gaussian", they lead to different joint probabilistic models $p(\y,\x,\s)$ and  therefore to different methodologies. 

For a given BG model, the algorithms of the literature differ in the choice of the optimization criterion and the practical implementation they consider. 
Contributions \cite{Larsson_ieeetsp07,Schniter_itaw08,Zayyani_icassp08,Baron_arxiv09,Herzet_icassp10} are based on model \eqref{eq:obsprob2}-\eqref{eq:priorsi2}. 
In \cite{Larsson_ieeetsp07}, the authors attempt to compute an (approximate) MMSE estimate of $\x$. To do so, they propose a heuristic procedure to identify the set of supports having the largest posterior probabilities $p(\s\vert\y)$.
 In \cite{Schniter_itaw08}, the authors derived the so-called ``Fast Bayesian Matching Pursuit" (FBMP) following a very similar approach. This contribution essentially differs from \cite{Larsson_ieeetsp07} in the way the set of supports with the largest posterior probabilities $p(\s\vert\y)$ is selected. 
  The approach considered in \cite{Zayyani_icassp08} is based on the joint maximization of $p(\y,\x,\s)$ for the decoding of real-field codes. More particularly, the authors consider a relaxation of the Bernoulli distribution $p(\s)$ and apply optimization techniques for smooth functions. 
The use of the sum-product algorithm \cite{Kschischang2001Factor} was investigated in \cite{Baron_arxiv09} to compute approximation of the marginals $p(\xs_i\vert \y)$. In the same spirit, another approach based on a mean-field approximation and the VB-EM algorithm \cite{Beal2003Variational}, has been considered in  \cite{Herzet_icassp10} to derive approximate values of $p(\x\vert \y)$, $p(\s\vert \y)$ and $p(\xs_i, \xs_i\vert \y)$. These approximate marginals are then  used to make approximate MMSE or MAP decisions on $\x$ and $\s$.  Finally, Ge \etal suggest in  \cite{Ge2011Enhanced} another approximation of $p(\x,\s | \y)$ based on a MCMC inference scheme.

On the other hand, model \eqref{eq:obsprob}-\eqref{eq:priorsi} has been  considered in \cite{Kormylo:1982uq,Soussen2011From, Dremeau_ssp11, Zayyani_icassp09}. Contribution \cite{Kormylo:1982uq} is the most related to the present work (in particular to BMP and BOMP): the authors proposed an ascent implementation of  two MAP problems,  involving  either $p(\y,\s)$ or $p(\y,\x,\s)$.  However, the subsets of variables over which the objective is maximized at each iteration differ from those considered in the implementation of the proposed BMP and BOMP. In \cite{Kormylo:1982uq}, the authors focus on a particular application, namely the denoising of ``geophysical signal" expressed in a wavelet basis, leading to a non-overcomplete setting. 
An extension to overcomplete settings has been considered by Soussen \etal in \cite{Soussen2011From}. In \cite{Dremeau_ssp11}, the authors focus on a MAP problem involving $p(\ss_i\vert\y)$ to make a decision on the support of the sparse vector. The intractability of the MAP problem is addressed by means of a mean-field variational approximation. Finally, a different approach is considered  in \cite{Zayyani_icassp09}: the authors make a decision on $\s$ by building a sequence of test of hypotheses; no particular  Bayesian objective function is considered and the construction of the tests are, to some extent, clever but heuristic.


%

\subsection{Connections with some well-known pursuit algorithms}

In this section, we emphasize the connections existing between the procedures derived in section \ref{sec:Bpursuit} and some well-known standard pursuit procedures. In particular, we show that BMP, BOMP, BStOMP and BSP can be seen as forward/backward extension of MP, OMP, StOMP and SP for some particular values of the parameters of  model \eqref{eq:obsprob}-\eqref{eq:priorsi}. 

We first show the equivalence of MP and BMP under the following setting: $\sigma^2_w\rightarrow0$, $\sigma^2_x \rightarrow \infty$.
Under these conditions, BMP updates \eqref{eq:explicitBMP1table}-\eqref{eq:explicitBMP2table} can be rewritten as:
\begin{align}
		\tilde{s}_i^{(n+1)}&=
		\left\{
		\begin{array}{ll}
		1 & \mbox{if $\langle \r^{(n)}+\hat{x}_i^{(n)} \d_i, \d_i \rangle^2 >0$,}\\
		0 & \mbox{otherwise,}
		\end{array}
		\right. \label{eq:particularBMP1}\\
		\tilde{x}_i^{(n+1)}&= \tilde{s}_j^{(n+1)}\left(\hat{x}_i^{(n)}+ \langle \r^{(n)}, \d_i \rangle\right).\label{eq:particularBMP2}
		\end{align}
First, note that any trace of $\sigma^2_x$ has disappeared in the resulting expressions; this intuitively makes sense since $\sigma^2_x \rightarrow \infty$ corresponds to a non-informative prior on $\x$. More importantly, the inequality in the right-hand side of \eqref{eq:particularBMP1} is then ``almost always" satisfied and therefore  $\tilde{s}_i^{(n+1)}=1$ $\forall i$. Indeed, $\tilde{s}_i^{(n+1)}=0$ occurs if and only if 
\begin{align}
\hat{x}_i^{(n)}= -\langle \r^{(n)}, \d_i \rangle. \label{eq:particularBMP3} 
\end{align}
Now, assuming that $\langle \r^{(n)}, \d_i \rangle$ is a continuously-valued random variable (which usually makes sense in practice, especially in noisy scenarios), we have that \eqref{eq:particularBMP3} is satisfied with probability zero. Assuming then that $\tilde{s}_i^{(n+1)}=1$ $\forall i$ and plugging \eqref{eq:particularBMP2} into \eqref{eq:residualidef}, we obtain 
\begin{align}
\rho^{(n)}(\tilde{x}^{(n+1)}_i,\tilde{s}^{(n+1)}_i)=\langle \r^{(n)}, \d_i \rangle^2 -\| \r^{(n)}\|^2. \label{eq:particularBMP4}
\end{align}
Finally, considering \eqref{eq:particularBMP2} and \eqref{eq:particularBMP4}, BMP recursions can be summarized as 
\begin{align}
j&=\argmax_i \langle \r^{(n)}, \d_i \rangle^2, \label{eq:particularBMP5}\\
\label{eq:particularBMP6}
	\hat{x}_i^{(n+1)}&=
	\left\{
	\begin{array}{ll}
	 \hat{x}_i^{(n)}+ \langle \r^{(n)}, \d_i \rangle&    \mbox{if $i=j$,} \\
	 \hat{x}_i^{(n)} &  \mbox{otherwise.} 
	\end{array}
	\right.
\end{align}
Now, recursions \eqref{eq:particularBMP5}-\eqref{eq:particularBMP6} exactly correspond to the definition of MP \cite{Mallat_ieeetsp93}. Hence, in the particular case where $\sigma^2_w\rightarrow0$ and $\sigma^2_x \rightarrow \infty$, BMP turns out to be equivalent to MP. In the general case, however, the two algorithms significantly differ since BMP implements features that are not available in MP. In particular, BMP implements atom deselection as soon as $\sigma^2_w \neq 0$ whereas MP only allows for atom selection. Moreover, unlike MP, BMP can take into account some information about the atom occurrence ($p_i$'s) or the variance of the non-zero coefficients ($\sigma^2_x$). 

Similarly, OMP turns out to be a particular case of BOMP under the conditions $\sigma^2_w\rightarrow0$, $\sigma^2_x \rightarrow \infty$. 
 Indeed, first remind that the first step of  BOMP \eqref{eq:mainUpdateBOMP1} is strictly equivalent to BMP update \eqref{eq:mainUpdateBMP1}-\eqref{eq:mainUpdateBMP2}. Hence, from the discussion above we have that BOMP support update can be rewritten as
 \begin{align}
\label{eq:particularBOMP6}
	\hat{s}_i^{(n+1)}&=
	\left\{
	\begin{array}{ll}
	 1&    \mbox{if $i=j$,} \\
	 \hat{s}_i^{(n)} &  \mbox{otherwise,} 
	\end{array}
	\right.
\end{align}
where $j$ is defined in \eqref{eq:particularBMP5}. 
Moreover, we have that
\begin{align}
\lim_{\sigma^2_w\rightarrow 0}  \left(\D^T_{\s} \D_{\s}+ \frac{\vars_w}{\vars_x}\I_{\| \s \|_0} \right)^{-1}   \D^T_{\s} \y = \D^\dag_{\s}\y.
\end{align}
Hence, BOMP update \eqref{eq:BOMPcoefupdatetable} becomes
\begin{align}
\hxp_{\hat{\s}^{(n+1)}}&=\D^\dag_{\hat{\s}^{(n+1)}}\y. \label{eq:particularBOMP7}
\end{align}
Now, \eqref{eq:particularBMP5}, \eqref{eq:particularBOMP6} and \eqref{eq:particularBOMP7} correspond to the standard implementation of OMP \cite{Pati_asilomar93}.

Let us finally compare the recursions implemented by StOMP and BStOMP in the particular case where $\sigma^2_x\rightarrow \infty$. 
Since BOMP and BStOMP (resp. OMP and StOMP) implement the same coefficient update, we only focus on the update of the support vector $\s$. 
First let us remind that StOMP support update is expressed as
\begin{align} 
\hat{s}_i^{(n+1)}&=
\left\lbrace\label{eq:thresholdBStOMPequivStOMP}
\begin{array}{ll}
1 & \mbox{if $\langle \r^{(n)}, \d_i \rangle^2 > \tilde{T}^{(n)}$,}\\
\hat{s}_i^{(n)} & \mbox{otherwise,}
\end{array}
\right.
\end{align} 
where $\tilde{T}^{(n)}$ is a threshold which derives from hypothesis-testing considerations \cite{Donoho_StOMP06}. It is clear from \eqref{eq:thresholdBStOMPequivStOMP} that StOMP is a forward algorithm; in particular, it selects at each iteration all atoms whose correlation with current residual exceeds a certain threshold. 

On the other hand, if $\sigma^2_x \rightarrow \infty$, BStOMP support update \eqref{eq:BStOMPcoefupdatetable1} can be rewritten as
\begin{align} 
\hat{s}_i^{(n+1)}&=
\left\lbrace\label{eq:thresholdStOMP}
\begin{array}{ll}
1 & \mbox{if $\langle \r^{(n)}+\hat{x}_i^{(n)} \d_i, \d_i \rangle^2 > T_i $},\\
0 & \mbox{otherwise.}
\end{array}
\right.
\end{align} 
In particular,  if the $i$th atom was not selected at iteration $n-1$, \ie $(\hat{x}_j^{(n)}, \hat{s}_j^{(n)})=(0,0)$, this expression becomes
\begin{align} 
\hat{s}_i^{(n+1)}&=
\left\lbrace\label{eq:thresholdBStOMPequivStOMP2}
\begin{array}{ll}
1 & \mbox{if $\langle \r^{(n)}, \d_i \rangle^2 > T_i$,}\\
\hat{s}_i^{(n)} & \mbox{otherwise.}
\end{array}
\right.
\end{align} 
Comparing \eqref{eq:thresholdBStOMPequivStOMP2} to \eqref{eq:thresholdBStOMPequivStOMP}, we see that the support update of StOMP and BStOMP are similar in such a case. However, in the general case \eqref{eq:thresholdStOMP}, BStOMP allows for the deselection of atoms. Moreover, another crucial difference between StOMP and BStOMP lies in the definition of the threshold $T_i$ and $\tilde{T}^{(n)}$. Indeed, the Bayesian framework considered in this paper naturally leads to a definition of the threshold as a function of the model parameters. Unlike the approach followed in \cite{Donoho_StOMP06}, it requires therefore no additional hypothesis and/or design criterion. 

Finally, BSP reduces to SP as $\sigma^2_w\rightarrow0$, $\sigma^2_x \rightarrow \infty$ and for certain modifications of the sets $\Sc_P$ and $\Sc_K$ appearing in \eqref{eq:BMPsu1} and \eqref{eq:BMPsu2}. The proof of this connection is slightly more involved to establish and is reported to  Appendix C.

\section{Simulation Results }
\label{sec:results}

In this section we illustrate the performance of the proposed algorithms. We evaluate the following metrics by Monte-Carlo simulation: \emph{i)} the mean square error (MSE) on the non-zero coefficients of the sparse vector; \emph{ii)}
the probability of wrong decision on the support, \ie $P_e \triangleq p(\hat{s}_i \neq s_i)$. For each point of simulation, we averaged the results over $3000$ trials. 

%
%

\subsection{The Uniform Case}

We first consider the case where all the atoms have the same probability to be active, \ie $p_i=p\ \forall i.$ For each experiment, the data vector $\y$ is generated according to model \eqref{eq:obsprob} with $\sigma_n^2=10^{-4}$ and $\sigma_x^2=1$.
In Fig. \ref{fig:uniformvsK} we represent the MSE, the probability of wrong decision on the elements of the support and the average running time achieved by different sparse-representation algorithms. Each point of simulation corresponds to a fixed number of non-zero coefficients, say $K$, and, given this number, the positions of the non-zero coefficients are drawn uniformly at random for each observation. We set $N=154$, $M=256$. 

In addition to the proposed procedures, we consider several algorithms of the state of the art: MP \cite{Mallat_ieeetsp93}, OMP \cite{Pati_asilomar93}, StOMP \cite{Donoho_StOMP06}, SP \cite{Dai_arxiv09}, IHT \cite{Blumensath_jfaa08}, HTP \cite{Foucart_tr11}, Basis Pursuit Denoising (BPD) \cite{Chen_siam99}, SBR \cite{Soussen2011From}, SOBAP \cite{Dremeau_ssp11} and FBMP \cite{Schniter_itaw08}. 
 The stopping criterion used for MP and OMP is based on the norm of the residual: the recursions are stopped as soon as the norm of the residual drops below $\sqrt{N\sigma_n^2}$. StOMP is run with the ``CFAR" thresholding criterion \cite{Donoho_StOMP06}. BStOMP and BSP implement the noise variance estimation described in section \ref{sec:BayEfroymson}. We set $p_i=\frac{K}{M}, \; \forall i$ in all the Bayesian pursuit algorithms. 

We see in Fig. \ref{fig:uniformvsK} that the Bayesian pursuit algorithms outperform their standard counterparts (MP, OMP, StOMP and SP) both in terms of MSE and probability of wrong detection of the support.  The performance improvement depends upon the considered algorithms: whereas the gain brought by BMP, BStOMP and BSP is significant, BOMP only leads to a marginal improvement in terms of MSE.  We also notice that the running times of the Bayesian and standard procedures are roughly similar. 

The proposed Bayesian pursuit algorithms also perform well with respect to other algorithms of the literature. In particular, BSP and BStOMP are only outperformed by FBMP and SOBAP (resp. SOBAP) in terms of MSE (resp. probability of error on the support). The gain in performance allowed by FBMP and SOBAP  has however to be contrasted with their complexity since they involve a much larger computational time.

We repeated similar experiments for many different values of $N$ and $K$  with $M=256$, and we came to similar conlusions.  
Fig. \ref{fig:PTuniform} illustrates the results of these experiments in a ``phase diagram": each curve represents the couples $(K/N,N/M)$ for which a particular algorithm achieves MSE=$10^{-2}$ (top) or $P_e=10^{-2}$ (bottom). Above (resp. below) these curves, the corresponding algorithms perform worse (resp. better). For the sake of readibility, we only reproduce the results for the proposed procedures and some of the algorithms considered in  Fig. \ref{fig:uniformvsK}. 
 As previously, we see that the Bayesian pursuit algorithms improve the performance with respect to their standard counterparts. The gain  is the most significant for moderate-to-high value of $N/M$. We note the bad behavior of StOMP in terms of probability of error on the support: the algorithm achieved $P_e=10^{-2}$ for no value of the parameters $(K/N,N/M)$. This is due to the fact that StOMP always selects too many atoms and can never removed them at subsequent iterations. In contrast, we see that the atom deselection process implemented by BStOMP solves this problem and leads to very good results in terms of support recovery. 
 
 Finally, we also assess the robustness of the proposed approaches to different levels of noise. Once again, we have observed that the conclusions drawn previously remain valid at lower signal-to-noise ratios. Fig. \ref{fig:MSEvsNoise}, which represents the MSE versus the noise variance, illustrates our purpose. We see that the Bayesian pursuit algorithms allow for an improvement of the performance irrespective of the noise level. The most effective procedures remain the more complex SoBaP and FBMP.

\subsection{The Non-uniform Case}

We now consider the case where the atoms of the dictionary have different probabilities to be active. 
 We assume that the $p_i$'s are independent realizations of a beta distribution $Beta(\alpha,\beta)$ with $\alpha=0.4$, $\beta=0.4$. The data are generated as  previously. In particular, for each trial, the positions of the non-zero coefficients are drawn uniformly at random. This leads to a realization of $\s$. Knowing $\s$, we draw the value of $p_i$ from its posterior distribution\footnote{It is easy to show that $p(p_i\vert s_i)$ is a beta distribution $Beta(\alpha',\beta')$ with $\alpha'=\alpha+s_i$ and $\beta'=\beta+1-s_i$.} $p(p_i\vert s_i)$ $\forall\, i$.   

In Fig. \ref{fig:NonUniformvsK}, we represent the performance of the Bayesian pursuit algorithms when they have access to the values of the occurrence probabilities $p_i$'s. For the sake of comparison, we use the same simulation parameters as those used to generate Fig. \ref{fig:uniformvsK}. The red curves reproduce the performance of the Bayesian pursuit algorithms observed in Fig. \ref{fig:uniformvsK}. 
 The blue curves illustrate the performance of the same algorithms when they are fed with the values of $p_i$'s.  We also show the algorithm performance when they are fed with noisy estimates, say $\hat{p}_i$'s, of the actual prior $p_i$ (green curves) :
 \begin{align}
\hat{p}_i= p_i + \mathcal{U}(\max(0,p_i-\Delta_p), \min(1,p_i+\Delta_p)),
\end{align}
where $\mathcal{U}(a,b)$ denotes a uniform on $[a,b]$ and $\Delta_p$ is a parameter determining the level of the perturbation. We set $\Delta_p=0.3$ in Fig. \ref{fig:NonUniformvsK}. 

Since the $p_i$'s constitute an additional source of information about the position of the non-zero coefficients in the sparse vector,  the Bayesian pursuit algorithms are expected to improve the recovery performance. This is observed in Fig. \ref{fig:NonUniformvsK}: all the Bayesian algorithms achieve better performance when informed of the values of $p_i$'s.  The improvement of  the performance of BMP and BOMP is only marginal. On the contrary, a significant gain of performance is observed for BStOMP and BSP, both in terms of MSE and probability of error on the support. We can also see that the algorithms are quite robust to an error on the a priori probabilities: the setup $\Delta_p=0.3$ degrades but still allows improvements with respect to the non-informed case. 

As in the uniform case, we repeated these experiments for many different values of $K$ and $N$, with $M=256$. This leads to the phase diagrams represented in Fig. \ref{fig:PTnonuniform}. Each curve represents the set of couples $(K/N,N/M)$ for which a particular algorithm achieves MSE=$10^{-2}$ (top) or $P_e=10^{-2}$ (bottom). The curves obtained in Fig. \ref{fig:PTuniform} for the uniform case (that is using  $p_i=\frac{K}{M}, \; \forall i$ in the algorithms) are reproduced in red. The blue curves illustrate the performance of the algorithms informed with the values of $p_i$'s.  We see that the conclusions drawn from Fig. \ref{fig:NonUniformvsK} are confirmed by the phase diagram: the exploitation of the prior information enables to enhance the algorithm performance. The most impressive improvements are obtained by BSP and BStOMP. We also note the good behavior of BOMP for large values of $N/M$.

\section{Conclusions}

The contribution of this paper is three-fold. First, we emphasized a connection between the standard penalized sparse problem and a MAP problem involving a BG model. Although this model has been previously involved in several contributions, its link with the standard sparse representation problem had been poorly addressed so far. Only recently, some contributions proposed results in that vein. The result derived in this paper is however more general since it requires weaker conditions on the observed vector $\y$. 

Secondly, we proposed several methodologies to search for the solution of the considered MAP problem. The proposed algorithms turn out to be forward/backward pursuit algorithms, which account for the a priori information available on the atoms occurrences. We showed on different experimental setups the good behavior of the proposed procedures with respect to several algorithms of the literature. 

Finally, we established interesting connections between the proposed procedures and well-known algorithms of the literature: MP, OMP, StOMP and SP. We believe that these connections build enlightening bridges between these standard procedures (which have been known, for some of them, for more than two decades now) and Bayesian procedures which have attracted the attention of the sparse-representation community more recently. 

In our future work, we will focus on the derivation of guarantees for the recovery of the ``true" support of the sparse decomposition. To the best of our knowledge, the derivation of such guarantees for forward/backward algorithms mostly remain an open (and difficult) problem in the literature. A few results have been proposed in particular settings, see \eg \cite{Needel_ACMels08, Dai_arxiv09}. However, no general guarantees exists for many (even very simple) forward/backward algorithms as \cite{Efroymson_60, Berk1980Forward, Broersen_jrssc86}.

\appendices

\section{Proof of Theorem 1}\label{sec:Proof_Th1}

In this appendix, we prove the equivalence between \eqref{eq:slsrp} and \eqref{eq:BGp} under the hypotheses of Theorem 1, \ie 
\textit{i)} $\| \y \|_2 < \infty$, 
\textit{ii)} $\vars_{x}\rightarrow \infty$,
\textit{iii)} $p_i=p$ $\forall i$,
\textit{iv)} $\lambda\triangleq 2\vars_w \log(\frac{1-p}{p})$. 
 For the sake of clarity and conciseness, we restrict the demonstration to the case where any subset of $L\leq N$ columns of $\D$ is linearly independent. The general case can however be derived in a similar way.

We first pose some notations and definitions. 
Let 
\begin{align}
\Xc^\star &\triangleq \arg\min_\x  \{\|\y-\D\x \|^2_2+\lambda \| \x\|_0 \},\nonumber
\end{align}
be the set of solutions of the standard sparse representation problem.
We define $f(\s)$ as
\begin{align} \label{eq:stdproA1FP} 
f(\s) \triangleq
& \min_{\x} \{\|\y-\D\x \|^2_2+\lambda \| \x\|_0 \}\\
& \mbox{ s.t. $x_i=0 \;$ if $s_i=0$}.\nonumber
\end{align}
$f(\s)$ is therefore the minimum of the standard SR problem \eqref{eq:slsrp} if the support of the sparse vector is fixed. Moreover we define $\x^\star(\s)$ as
\begin{align}\label{eq:xstar1FP} 
\x_\s^\star(\s)\triangleq\D_\s^\dag \y, \mbox{ and }
x_i^\star(\s)\triangleq 0  \mbox{ if $s_i=0$.}
\end{align}
Clearly, $\x^\star(\s)$ is a solution of \eqref{eq:stdproA1FP}. 
It is the unique (resp. the minimum $\ell_2$-norm) solution if $\| \s \|_0 \leq N$ (resp. $\| \s \|_0 > N$). Using these notations, we can redefine $\Xc^\star$ as follows 
\begin{align}
\Xc^\star &= \{ \x \in \R^M \vert \x=\x^\star(\s)\, \mbox{ with $\s\in\argmin_\s f(\s)$} \}.\label{eq:defXcstarFP} 
\end{align}
This definition is valid because the minimum of $f(\s)$ is necessarily achieved for $\s$ such that $\| \s\|_0\leq N$, in which case \eqref{eq:xstar1FP}  is the unique solution of  \eqref{eq:stdproA1FP}.  

Let moreover 
\begin{align}
g(\s,\sigma_x^2) &\triangleq 2 \vars_w \min_{\x} \{- \log p(\y,\x,\s) - \log(1-p)\}, \label{eq:BGsrp1FP}\\
g(\s) &\triangleq \lim_{\sigma_x^2\rightarrow \infty} g(\s,\sigma_x^2),\\
\hat{\x}(\s)&\triangleq \argmin_{\x} \{-\log p(\y,\x,\s)\}. \label{eq:BGsrp2FP}
\end{align}
The goal function in \eqref{eq:BGsrp1FP} is equal, up to an additive constant, to $-\log p(\y,\x,\s)$ and $g(\s)$ corresponds to its minimum (over $\x$) when $\sigma^2_x\rightarrow \infty$. 
The equality in \eqref{eq:BGsrp2FP} is well-defined since the  minimum exists and is unique as shown below. With these notations, we can define the set of solutions of the BG problem \eqref{eq:BGp} when $\sigma_x^2\rightarrow \infty$ as
\begin{align}\label{eq:def_hatXcFP}
\hat{\Xc} \triangleq \{ \x \in \R^M \vert \x=\hat{\x}(\s)\, \mbox{ with $\s\in\argmin_\s g(\s)$} \}.
\end{align}
We want therefore to show that $\Xc^\star = \hat{\Xc}$ under the hypotheses of Theorem 1.


The proof of Theorem 1 is based on the following intermediate results:\vspace{-0.2cm}
\begin{align}
\lim_{\vars_{x}\rightarrow \infty} \hat{\x}(\s) &= \x^\star(\s)  \quad \forall\s,\label{eq:intres1FP}\\
g(\s)& \geq f(\s)\quad\ \forall \s,\label{eq:intres2FP} \\
\{ f(\s) \vert\, \s\in \{ 0, 1\}^M \} &\subseteq \{ g(\s) \vert\, \s\in \{ 0, 1\}^M \},\label{eq:intres2bFP}\\
\min_{\s} g(\s)&=\min_{\s} f(\s),\label{eq:equalityminFP}\\
 \argmin_\s g(\s) &\subseteq \argmin_\s f(\s) \label{eq:intres3FP}.
\end{align}

\subsubsection{Proof of \eqref{eq:intres1FP}} 
From standard Bayesian theory (see for example \cite[Chap. 14]{Mendel_book}), it can be seen that the unique minimum of $\log p(\y,\x,\s)$ is given by
\begin{align}\label{eq:xhat1}
\begin{array}{ll}
\hat{\x}_\s(\s) = \left( \D_\s^T \D_\s+ \frac{\vars_w}{\vars_x}\I_k \right)^{-1} \D_\s^T \y, & \\
\hat{x}_i(\s)=0 & \mbox{if $s_i=0$.}
\end{array}
\end{align}
 Taking the limit for $\vars_{x}\rightarrow \infty$, we obtain the equivalence between \eqref{eq:xstar1FP} and \eqref{eq:xhat1}.\vspace{0.3cm}

\subsubsection{Proof of \eqref{eq:intres2FP}} 
Using \eqref{eq:intres1FP} and taking \eqref{eq:obsprob}-\eqref{eq:priorsi} into account, we have
\begin{align} 
g(\s)&=
\| \y -\D \x^\star(\s)\|^2_2- 2\vars_w \log p(\s)
- 2\vars_w \log(1-p)\nonumber \\
& + \vars_w \lim_{\vars_x\rightarrow\infty} \frac{ \| \hat{\x}(\s)\|_2^2}{\vars_x}.\label{eq:res2eq1}
\end{align}
Since $\| \y \|_2<\infty$ by hypothesis, it follows that 
\begin{align}
\lim_{\vars_x\rightarrow\infty}\| \hat{\x}(\s)\|_2=  \| \x^\star(\s)\|_2=\|\D_\s^\dag \y \|_2<\infty,\vspace{-0.3cm}
\end{align}
since $\D_\s^\dag$ is a bounded operator. Hence, the last term in \eqref{eq:res2eq1} tends to zero.  Moreover,  since $p_i=p$ $\forall i$ one can rewrite $p(\s)$ as 
 \begin{align}
\log p(\s)= - \| \s \|_0 \log\left(\frac{1-p}{p}\right) + \log(1-p). 
\end{align}
Therefore, letting $\lambda\triangleq 2\vars_w \log(\frac{1-p}{p})$, we obtain
\begin{align}
g(\s)&=
\| \y -\D \x^\star(\s)\|^2_2+ \lambda \| \s \|_0. 
\end{align}
Finally, realizing that 
\begin{align}
\| \x^\star(\s)\|_0\leq \| \s\|_0\quad \forall \s,\label{eq:needalabelhere}
\end{align}
 we come up with \eqref{eq:intres2FP}.   It is interesting to note that the equality holds in \eqref{eq:intres2FP} if and only if  $\| \x^\star(\s)\|_0=\| \s\|_0$, \ie when $x_i^\star(\s) \neq 0$ $\forall s_i=1$.\vspace{0.3cm}

\subsubsection{Proof of \eqref{eq:intres2bFP}} 
To prove this result, we show that $\forall \s\in \{0,1\}^M$, $\exists\, \tilde{\s}\in \{0,1\}^M$ such that $f(\s)= g(\tilde{\s})$.  First, notice that the value of $f(\s)$ is only a function of $\x^\star(\s)$. Now, $\forall \s$ one can find $\tilde{\s}$ such that 
\begin{align} 
\x^\star(\s)&=\x^\star(\tilde{\s}),\label{eq:ir2p1e1}\\
\| \x^\star(\tilde{\s})\|_0 &= \| \tilde{\s} \|_0 \label{eq:ir2p1e2}. 
\end{align}
In order to see the last assertion, define $\tilde{\s}$ as
\begin{align}\label{eq:defstilde}
\tilde{s}_i&= \left\{
\begin{array}{ll}
1 & \mbox{if $x_i^\star(\s)\neq 0$}\\
0 & \mbox{otherwise}.
\end{array}\right.
\end{align}
Clearly, $\s$ and $\tilde{\s}$ differ as soon as $\| \x^\star(\s) \|_0\neq \| \s \|_0$. Considering problem \eqref{eq:stdproA1FP}, $\s$ and $\tilde{\s}$ introduce therefore different sets of constraints on the solution. By definition, the contraints defined by $\tilde{\s}$ include those defined by $\s$. However, the new constraints introduced by $\tilde{\s}$ has no effect on the solution since they correspond to  $x_i^\star(\s)=0$. Then clearly, \eqref{eq:ir2p1e1} and \eqref{eq:ir2p1e2} follow.
Finally, \eqref{eq:ir2p1e1} ensures that $f(\s)=f(\tilde{\s})$ and \eqref{eq:ir2p1e2} implies $f(\tilde{\s})= g(\tilde{\s})$ as emphasized in the remark below \eqref{eq:needalabelhere}. This shows \eqref{eq:intres2bFP}. \vspace{0.3cm}

%
%

\subsubsection{Proof of \eqref{eq:equalityminFP} and \eqref{eq:intres3FP}} 
Let $\s^\star$ and $\hat{\s}$ be minimizers of $f(\s)$ and $g(\s)$ respectively. Then, we have
\begin{align}
f(\s^\star)&\stackrel{(a)}{=} g(\tilde{\s})\ \mbox{for some $\tilde{\s}$,} \nonumber \\
&\stackrel{(b)}{\geq} g(\hat{\s}) \stackrel{(c)}{\geq} f(\hat{\s}) \stackrel{(d)}{\geq} f(\s^\star) ,\label{eq:totineq}
\end{align}
where \emph{(a)} is a consequence of \eqref{eq:intres2bFP};  \emph{(c)} results from \eqref{eq:intres2FP} and \emph{(b)}, \emph{(d)} follow from the definition of $\s^\star$ and $\hat{\s}$. Looking at the right and left-most terms, we see that the equality holds throughout \eqref{eq:totineq}. Hence, 
\eqref{eq:equalityminFP} results from $f(\s^\star)=g(\hat{\s})$ and \eqref{eq:intres3FP} from $f(\s^\star)=f(\hat{\s})$.

\subsubsection{Proof of $\hat{\Xc}= \Xc^\star$} 
First, it is easy to see that the inclusion $\hat{\Xc}\subseteq \Xc^\star$ holds by considering the definition of $\Xc^\star$ and $\hat{\Xc}$ (in \eqref{eq:defXcstarFP} and  \eqref{eq:def_hatXcFP}) and the technical results \eqref{eq:intres1FP} and \eqref{eq:intres3FP}.

The reverse inclusion ($\Xc^\star \subseteq \hat{\Xc}$) can be proved   by showing that $\forall\, \s^\star \in  \argmin_\s f(\s)$, $\exists\,  \hat{\s}\in \argmin_\s g(\s)$
 such that
\begin{align}
\x^\star(\s^\star)= \lim_{\vars_{x}\rightarrow \infty} \hat{\x}(\hat{\s}). \label{eq:secstatMR} \vspace{-1.0cm}
\end{align}
For any $\s^\star$, let us define $\hat{\s}$ as
\begin{align}
\hat{s}_i &=\left\{
\begin{array}{ll}
0 & \mbox{if $x_i^\star(\s^\star) = 0$}\\
1 & \mbox{otherwise}
\end{array}
\right.
\end{align}
and let us show that such an $\hat{\s}$ satisfies the above conditions. From this definition, we have that
 \begin{align}
\| \x^\star(\hat{\s})\|_0=\| \hat{\s}\|_0, \label{eq:hmmmmffffpsssscch}\\
 \x^\star(\s^\star)= \x^\star(\hat{\s}). \label{eq:egxstrxhat}
\end{align}
Hence, combining \eqref{eq:intres1FP} with the last equality, we obtain \eqref{eq:secstatMR}. It thus remains to show that $\hat{\s}$ is a minimizer of $g(\s)$. This can be seen from the following sequence of equalities: 
\begin{align}\label{eq:last?}
f(\s^\star)&\stackrel{(a)}{=}f(\hat{\s})\stackrel{(b)}{=} g(\hat{\s})\stackrel{(c)}{=} \min_{\s} g(\s), 
\end{align}
where \emph{(a)} follows from \eqref{eq:egxstrxhat}; \emph{(b)} is a consequence of \eqref{eq:hmmmmffffpsssscch}; and \emph{(c)} results from \eqref{eq:equalityminFP} and the definition of $\s^\star$.

\section{Derivation of Bayesian Pursuit Algorithms} \label{sec:derivBMP}

\subsection{BMP}
Let us define
\begin{align}
\hat{\x}_i^{(n)}&\triangleq [\hat{x}^{(n)}_1\,\ldots\,  \hat{x}^{(n)}_{i-1} \ x_i \  \hat{x}^{(n)}_{i+1}\,\ldots\, \hat{x}^{(n)}_M]^T,\\
\hat{\s}_i^{(n)}&\triangleq [\hat{s}^{(n)}_1\,\ldots\,  \hat{s}^{(n)}_{i-1} \ s_i \  \hat{s}^{(n)}_{i+1}\,\ldots\, \hat{s}^{(n)}_M]^T,
\end{align}
and
\begin{align}
\rho^{(n)}(x_i,s_i)\triangleq \sigma^2_w \log p(\y,\hat{\x}_i^{(n)},\hat{\s}_i^{(n)}).  \label{eq:defrhoanx-1}
\end{align}
Using the BG model defined in \ref{sec:connection}, we obtain
\begin{align}
\rho^{(n)}(x_i,s_i)\negmedspace
=&\sigma^2_w \negmedspace \left(\log p(\y \vert \hat{\x}_i^{(n)},\hat{\s}_i^{(n)}) \negmedspace+\negmedspace\log p(\hat{\x}_i^{(n)})\negmedspace+\negmedspace \log p(\hat{\s}_i^{(n)})\right)\negmedspace, \nonumber\\
\propto&  -\| \r^{(n)} +(\hat{s}_i^{(n)}\hat{x}_i^{(n)}-s_i x_i) \d_i \|^2
- \epsilon\, x_i^2 - \lambda_i s_i, \label{eq:defrhoanx0}
\end{align}
where $\epsilon=\sigma^2_w/\sigma^2_x$ and $\lambda_i=\sigma^2_w \log((1-p_i)/p_i)$.

Then, clearly \eqref{eq:mainUpdateBMP1}-\eqref{eq:mainUpdateBMP2} and \eqref{eq:mainUpdateBMP3} are tantamount to solving
\begin{align}
(\tilde{x}_i^{(n+1)},\tilde{s}_i^{(n+1)})
  &= \argmax_{(x_i,s_i)} \rho^{(n)}(x_i,s_i)\quad \forall\, i,\label{eq:equivalence1anx} \\
j &= \argmax_{i} \rho^{(n)}(\tilde{x}_i^{(n+1)},\tilde{s}_i^{(n+1)}) \label{eq:equivalence2anx},
\end{align}
and setting $\hat{s}_i^{(n+1)}$ and $\hat{x}_i^{(n+1)}$ as in \eqref{eq:BMPSRU} and \eqref{eq:BMPSCU}. 

Let us then derive the analytical expression  of the solution of \eqref{eq:equivalence1anx}.
First, we have
\begin{align}
\tilde{x}_i(s_i)
&\triangleq\argmax_{x_i} \rho^{(n)}(x_i,s_i),\nonumber\\
&= s_i \frac{\vars_x}{\vars_x+\vars_w}\left(\hat{x}_i^{(n)}+ \langle \r^{(n)}, \d_i \rangle\right).\label	{eq:explicitBMP2tableanx}
\end{align}
Indeed, since $\rho^{(n)}(x_i,s_i)$ is a convex function of $x_i$, the last equality is simply obtained by solving $\frac{\partial\rho^{(n)}(x_i,s_i)}{\partial x_i}=0$. Moreover, by definition, $\tilde{s}_i^{(n+1)}$ can be expressed as
\begin{align}
\tilde{s}_i^{(n+1)}&= \label{eq:defstildeianx}
\left\{
\begin{array}{ll}
 1 &    \mbox{if $\rho^{(n)}(\tilde{x}_i(1),1)>\rho^{(n)}(\tilde{x}_i(0),0)$,} \\
 0 &  \mbox{otherwise,} 
\end{array}
\right.
\end{align}
Using the definitions of $\rho^{(n)}(x_i,s_i)$ and $\tilde{x}_i(s_i)$, the inequality in the right-hand side of \eqref{eq:defstildeianx} is equivalent to 
\begin{align}
\langle \r^{(n)}+\hat{x}_i^{(n)} \d_i, \d_i \rangle^2 > 2\vars_w \frac{\vars_x+\vars_w}{\vars_x} \log\left(\frac{1-p_i}{p_i}\right). \label{eq:ineqanx1}
\end{align}
Combining, \eqref{eq:explicitBMP2tableanx}, \eqref{eq:defstildeianx} and \eqref{eq:ineqanx1}, we recover \eqref{eq:updaterestable}-\eqref{eq:thresholddeftable}.

\subsection{BOMP}

The first step of BOMP \eqref{eq:mainUpdateBOMP1} has been discussed in the previous subsection. We therefore focus on \eqref{eq:mainUpdateBOMP2}. We have that
\begin{align}
\log p(\y,\x,\hat{\s})\propto -\| \y - \D_{\hat{\s}} \x_{\hat{\s}} \|^2-\frac{\sigma^2_w}{\sigma^2_x} \| \x \|^2. \label{eq:BOMP1anx}
\end{align}
First, if $\hat{s}_i=0$, the corresponding element $x_i$ only appears in the second term in the right-hand side of \eqref{eq:BOMP1anx}, and $x_i=0$ clearly maximizes \eqref{eq:BOMP1anx}. It thus remains to find the value of $\x_{\hat{\s}}$. Since $\log p(\y,\x,\hat{\s})$ is a strictly concave function of $\x_\s$, the optimal value of $\x_\s$ can be found by solving $\nabla_{\x_\s} \log p(\y,\x,\hat{\s})= 0$ where
\begin{align}
\nabla_{\x_\s} \log p(\y,\x,\hat{\s}) = -2 \D_{\hat{\s}}^T \y - 2(\D_{\hat{\s}}^T \D_{\hat{\s}}+\frac{\sigma^2_w}{\sigma^2_x}\mathbf{I}) \, \x_{\hat{\s}}.\nonumber
\end{align}
Finally, noticing that $\D_{\hat{\s}}^T \D_{\hat{\s}}+\frac{\sigma^2_w}{\sigma^2_x}\mathbf{I}$ is definite positive (and therefore invertible), we obtain \eqref{eq:BOMPcoefupdatetable}.

\subsection{BSP}

The relation $\rho_i^{(n)}(s_i)= \rho^{(n)}(\tilde{x}_i(s_i),s_i)$ directly follows from \eqref{eq:defrhosi} and \eqref{eq:explicitBMP2tableanx}. 
Let us then elaborate on maximizations \eqref{eq:BMPsu1}-\eqref{eq:BMPsu2}.  The goal function of \eqref{eq:BMPsu1}-\eqref{eq:BMPsu2} can be expressed as\footnote{We drop the iteration index for conciseness.} 
\begin{align}
\sum_i \rho_i(s_i)
&\propto \sum_i \Delta(s_i)\triangleq h(\s),\nonumber
\end{align}
where $\Delta(s_i)\triangleq \rho_i(s_i)-\rho_i(\hat{s}_i^{(n)})$.
Let $\s$ be such that $s_i\neq\tilde{s}_i$ for $i \in \mathcal{I}$, where $\tilde{\s}$ is defined as in \eqref{eq:explicitBMP1table}; $h(\s)$ can be expressed as 
\begin{align}
h(\s) = \sum_i \Delta(\tilde{s}_i) - \left(\sum_{i \in \mathcal{I}} \Delta(\tilde{s}_i)-\Delta(1-\tilde{s}_i) \right)\negmedspace.
\end{align}
Now, by definition of $\tilde{s}_i$, we have $\Delta(\tilde{s}_i)\geq 0$, $\Delta(1-\tilde{s}_i)\leq 0$. Hence, the generic solution of \eqref{eq:BMPsu1}-\eqref{eq:BMPsu2} can be written as
\begin{align}
\hat{s}_i^{(n+1)}&=
\left\{
\begin{array}{ll}
1-\tilde{s}_i &\mbox{if $i\in\mathcal{I}^\star$},\\
\tilde{s}_i &\mbox{otherwise},
\end{array}
\right.
\end{align}
where
\begin{align}
\mathcal{I}^\star = \argmin_\mathcal{I} \sum_{i \in \mathcal{I}} \Delta(\tilde{s}_i)-\Delta(1-\tilde{s}_i),  \label{eq:anxBSPoptprob}
\end{align}
subject to some constraints on $\mathcal{I}$ depending on the considered problem. 
For instance, let us particularize the constraints on  $\mathcal{I}$ associated to \eqref{eq:BMPsu2}: if $\| \tilde{\s}\|_0\geq K$, then the definition of $\Sc_K$ implies that $\mathcal{I}$ must be such that 
\begin{align}\label{eq:ancCBSP1}
\begin{array}{l}
\tilde{s}_i=1\quad \quad\forall \, i\in \mathcal{I},
\\ \mathrm{Card}(\mathcal{I})=\| \tilde{\s}\|_0- K.
\end{array}
\end{align}
Similarly, if $\| \tilde{\s}\|_0< K$, the constraints become
\begin{align}\label{eq:ancCBSP2}
\begin{array}{ll}
\tilde{s}_i=0\quad \quad\forall \, i\in \mathcal{I},
\\ \mathrm{Card}(\mathcal{I})= K-\| \tilde{\s}\|_0. 
\end{array}
\end{align}
Hence, solving \eqref{eq:anxBSPoptprob} subject to \eqref{eq:ancCBSP1} (resp. \eqref{eq:ancCBSP2}) is equivalent to identifying the $\| \tilde{\s}\|_0- K$ (resp. $K-\| \tilde{\s}\|_0$) indices with $\tilde{s}_i=1$ (resp. $\tilde{s}_i=0$) leading to the smallest metrics $\Delta(\tilde{s}_i)-\Delta(1-\tilde{s}_i)$. 
Now, $\Delta(\tilde{s}_i)-\Delta(1-\tilde{s}_i)=\rho_i(\tilde{s}_i)-\rho_i(1-\tilde{s}_i)$ and $\rho_i(\ss_i)$ can be evaluated efficiently as described above. 

Problem \eqref{eq:BMPsu1} can be solved in a similar way and is not further described herefater.

\section{Connection between BSP and SP}

In this section, we emphasize the connection existing between BSP and SP when $\sigma^2_w\rightarrow 0$, $\sigma^2_x \rightarrow \infty$. 
For the sake of clarity, we first briefly remind SP implementation.  
SP operates in two steps. First, $P$  atoms are added to the previous support as follows:
\begin{align} 
\hat{\s}^{(n+\frac{1}{2})}= \argmax_{\s\in \tilde{\Sc}_P} \sum_i s_i \langle  \r^{(n)}, \d_i \rangle^2  \label{eq:SP2}
\end{align}
where 
\begin{align}
\tilde{\Sc}_P \triangleq \{ \s \vert\, \| \s-\hat{\s}^{(n)}\|_0=P, \mbox{$s_i=1$ if $s_i^{(n)}=1$ $\forall\,i$} \}.\label{defScPSP}
\end{align}
Then,  $\hat{\x}^{(n+\frac{1}{2})}$ is  evaluated according to \eqref{eq:particularBOMP7}.
 In a second step, SP removes some atoms from the support by applying the following rule:
\begin{align} 
\hat{\s}^{(n+1)}&= \argmax_{\s \in \tilde{\Sc}_K} \sum_i s_i\,  (\hat{x}_i^{(n+\frac{1}{2})})^2 \, \label{eq:SP1}
\end{align} 
where
\begin{align}
\tilde{\Sc}_K \triangleq \{ \s \vert\, \| \s\|_0=K, \mbox{$s_i=0$ if $\hat{s}_i^{(n+\frac{1}{2})}=0$ $\forall\,i$} \}.\label{defScKSP}
\end{align}
%
$\hat{\x}^{(n+1)}$ is evaluated from \eqref{eq:particularBOMP7}.

 We next show that BSP is equivalent to SP if we replace $\Sc_P$ by $\tilde{\Sc}_P$ in \eqref{eq:BMPsu1}, $\Sc_K$ by $\tilde{\Sc}_K$ in \eqref{eq:BMPsu2}, and let $\sigma^2_w\rightarrow 0$, $\sigma^2_x \rightarrow \infty$. We already showed the equivalence between the coefficient updates  \eqref{eq:BOMPcoefupdatetable} and \eqref{eq:particularBOMP7} in the previous section. It thus remains to prove the equivalence between \eqref{eq:SP2} and \eqref{eq:BMPsu1} (resp. \eqref{eq:SP1} and \eqref{eq:BMPsu2}). In order to do so, we use the following relation:
 \begin{align}\label{eq:anxusefulexpress}
\rho_i^{(n)}(1)-\rho_i^{(n)}(0)&=
\left\{
\begin{array}{ll}
\langle \r^{(n)} , \d_i\rangle^2 & \mbox{if $\hat{s}_i^{(n)}=0$}\\
-(\hat{\xs}_i^{(n)} )^2 & \mbox{if $\hat{s}_i^{(n)}=1$}
\end{array}
\right.
\end{align}
which can easily be proved by exploiting  \eqref{eq:explicitBMP2table} and \eqref{eq:residualidef}. Details are not reported here. 

Setting $\Sc_P=\tilde{\Sc}_P$, problem \eqref{eq:BMPsu1} can also be equivalently rewritten as
\begin{align} 
\hat{\s}^{(n+\frac{1}{2})}
&= \argmax_{\s\in\tilde{\Sc}_P} \sum_i (\rho_i^{(n)}(s_i)-\rho_i^{(n)}(0)),\nonumber\\
&= \argmax_{\s\in\tilde{\Sc}_P} \sum_{i} s_i \langle \r^{(n)} , \d_i\rangle^2 \label{eq:conBSP_SP1}
\end{align}
where the last equality follows from \eqref{eq:anxusefulexpress}. 
Similarly, setting $\Sc_K=\tilde{\Sc}_K$, problem \eqref{eq:BMPsu2} can be expressed as
\begin{align} 
\hat{\s}^{(n+1)}
&= \argmax_{\s\in\tilde{\Sc}_K} \sum_i (\rho_i^{(n)}(s_i)-\rho_i^{(n)}(1)),\nonumber\\
&= \argmax_{\s \in \tilde{\Sc}_K}  -\sum_i (1-s_i)\,  (\hat{x}_i^{(n+\frac{1}{2})})^2,\nonumber\\
&= \argmax_{\s \in \tilde{\Sc}_K}  \sum_i s_i\,  (\hat{x}_i^{(n+\frac{1}{2})})^2 \label{eq:conBSP_SP2}
\end{align}
where the last equalities are a consequence of \eqref{eq:anxusefulexpress}. Comparing \eqref{eq:conBSP_SP1} with \eqref{eq:SP2} (resp. \eqref{eq:conBSP_SP2} with \eqref{eq:SP1}) we obtain the result.

\singlespace
\bibliographystyle{IEEEbib}
\bibliography{group-15302,cherzet}

\twocolumn

\pgfplotsset{
every axis plot/.append style={mark size=2.5, line width=0.5pt},
every axis/.append style={font=\small},
label style={font=\small},
legend style={font=\tiny}
}

\begin{figure} 

\begin{tikzpicture}
\pgfplotstableread{Data_Simu_4/MSEa_dat/MSEa_rho0.6.dat}\loadedtable
\begin{semilogyaxis}[
xmax=100,
xlabel=$K$,
ylabel=Mean Square Error, 
grid=major,
legend columns=2,
legend style={at={(0.0,0.795)},anchor=west}
]
\addplot[color=red,mark=x,solid] table[x index=0, y index=1] {\loadedtable};
\addplot[color=blue,mark=x] table[x index=0, y index=5] {\loadedtable};
\addplot[color=red,mark=+,solid] table[x index=0, y index=2] {\loadedtable};
\addplot[color=blue,mark=+] table[x index=0, y index=6] {\loadedtable};
\addplot[color=red,mark=o,solid] table[x index=0, y index=3] {\loadedtable};
\addplot[color=blue,mark=o] table[x index=0, y index=7] {\loadedtable};
\addplot[color=red,mark=triangle,solid] table[x index=0, y index=4] {\loadedtable};
\addplot[color=blue,mark=triangle] table[x index=0, y index=8] {\loadedtable};
\addplot[color=black,mark=square] table[x index=0, y index=9] {\loadedtable};
\addplot[color=black,mark=star] table[x index=0, y index=10] {\loadedtable};
\addplot[color=cyan,mark=star] table[x index=0, y index=11] {\loadedtable};
\addplot[color=green,mark=star] table[x index=0, y index=12] {\loadedtable};
\addplot[color=green,mark=square] table[x index=0, y index=13] {\loadedtable};
\addplot[color=green,mark=o] table[x index=0, y index=14] {\loadedtable};
\legend{MP,BMP,OMP,BOMP,StOMP,BStOMP,SP,BSP,IHT,HTP,BPD,SOBAP,FBMP,SBR}

\end{semilogyaxis}
\end{tikzpicture}

\vspace{0.5cm}

\begin{tikzpicture}
\pgfplotstableread{Data_Simu_4/Pe_cw_dat/Pe_cw_rho_0.6.dat}\loadedtable
\begin{semilogyaxis}[
xmax=100,
xlabel=$K$,
ylabel=Probability of error on the support, 
grid=major,
legend columns=2,
legend style={at={(0.02,0.83)},anchor=west}]
\addplot[color=red,mark=x,solid] table[x index=0, y index=1] {\loadedtable};
\addplot[color=blue,mark=x] table[x index=0, y index=5] {\loadedtable};
\addplot[color=red,mark=+,solid] table[x index=0, y index=2] {\loadedtable};
\addplot[color=blue,mark=+] table[x index=0, y index=6] {\loadedtable};
\addplot[color=red,mark=o,solid] table[x index=0, y index=3] {\loadedtable};
\addplot[color=blue,mark=o] table[x index=0, y index=7] {\loadedtable};
\addplot[color=red,mark=triangle,solid] table[x index=0, y index=4] {\loadedtable};
\addplot[color=blue,mark=triangle] table[x index=0, y index=8] {\loadedtable};
\addplot[color=black,mark=square] table[x index=0, y index=9] {\loadedtable};
\addplot[color=black,mark=star] table[x index=0, y index=10] {\loadedtable};
\addplot[color=cyan,mark=star] table[x index=0, y index=11] {\loadedtable};
\addplot[color=green,mark=star] table[x index=0, y index=12] {\loadedtable};
\addplot[color=green,mark=square] table[x index=0, y index=13] {\loadedtable};
\addplot[color=green,mark=o] table[x index=0, y index=14] {\loadedtable};

\end{semilogyaxis}
\end{tikzpicture}

\vspace{0.5cm}

\begin{tikzpicture}
\pgfplotstableread{Data_Simu_4/Time_dat/Time_rho_0.6.dat}\loadedtable
\begin{semilogyaxis}[
xmax=100,
xlabel=$K$,
ylabel=Average Running Time (second), 
grid=minor,
legend columns=2,
legend style={at={(0.01,0.83)},anchor=west}]
\addplot[color=red,mark=x,solid] table[x index=0, y index=1] {\loadedtable};
\addplot[color=blue,mark=x] table[x index=0, y index=5] {\loadedtable};
\addplot[color=red,mark=+,solid] table[x index=0, y index=2] {\loadedtable};
\addplot[color=blue,mark=+] table[x index=0, y index=6] {\loadedtable};
\addplot[color=red,mark=o,solid] table[x index=0, y index=3] {\loadedtable};
\addplot[color=blue,mark=o] table[x index=0, y index=7] {\loadedtable};
\addplot[color=red,mark=triangle,solid] table[x index=0, y index=4] {\loadedtable};
\addplot[color=blue,mark=triangle] table[x index=0, y index=8] {\loadedtable};
\addplot[color=black,mark=square] table[x index=0, y index=9] {\loadedtable};
\addplot[color=black,mark=star] table[x index=0, y index=10] {\loadedtable};
\addplot[color=cyan,mark=star] table[x index=0, y index=11] {\loadedtable};
\addplot[color=green,mark=star] table[x index=0, y index=12] {\loadedtable};
\addplot[color=green,mark=square] table[x index=0, y index=13] {\loadedtable};
\addplot[color=green,mark=o] table[x index=0, y index=14] {\loadedtable};

\end{semilogyaxis}
\end{tikzpicture}
 \caption{MSE, probability of error on the support and average running time  versus the number of non-zero coefficients $K$ in the sparse vector.}\label{fig:uniformvsK}
\end{figure}
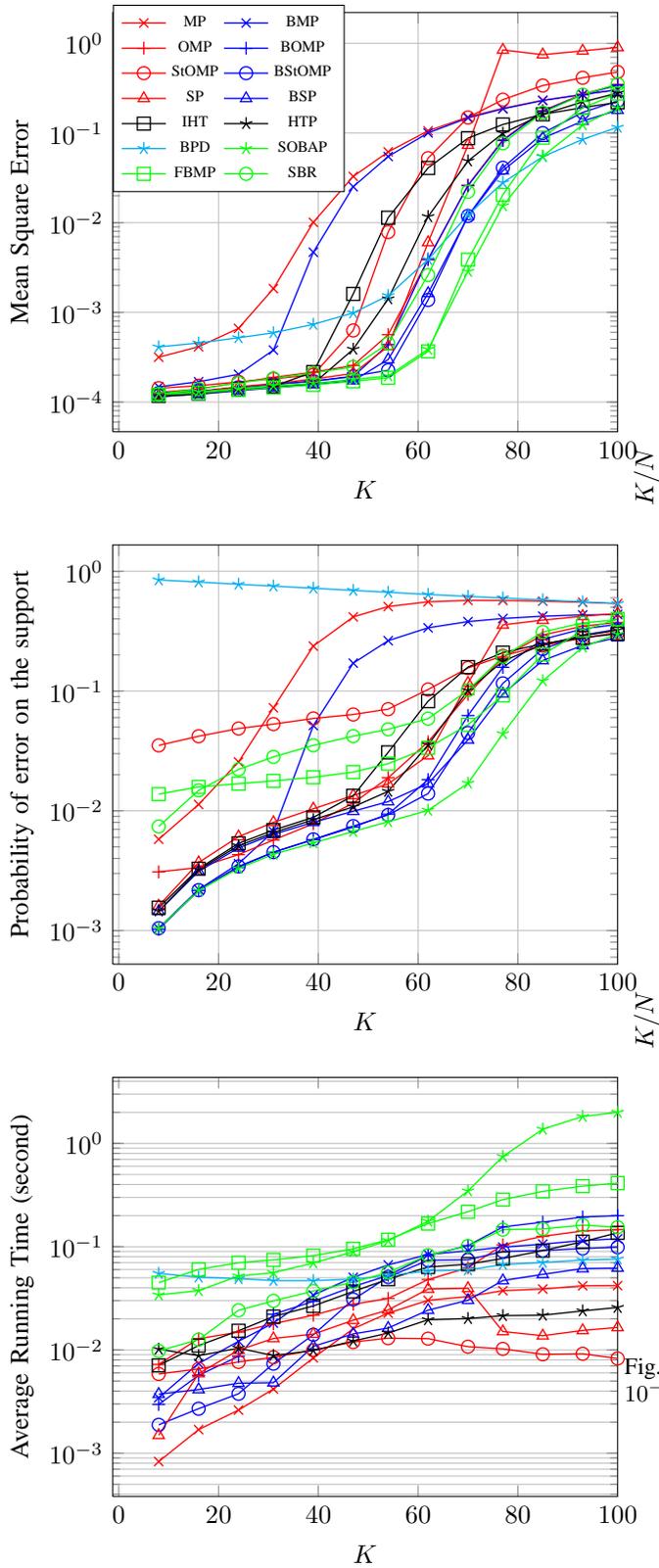

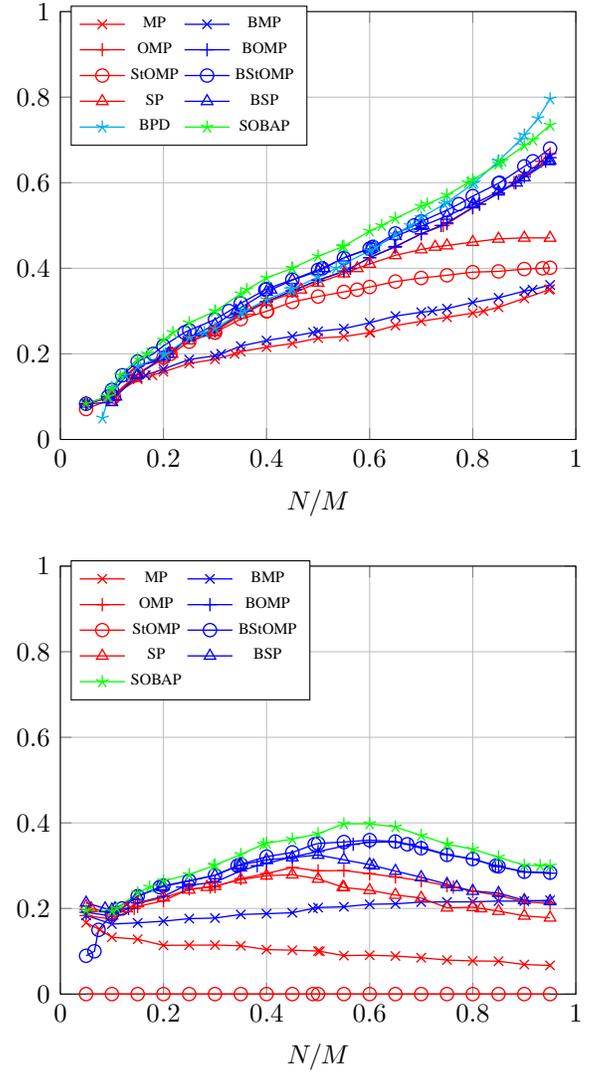
\begin{figure} 
\begin{tikzpicture}
\begin{axis}[
xmin=0,
xmax=1,
ymin=0,
ymax=1,
xlabel=$N/M$,
ylabel=$K/N$, 
grid=major,
legend columns=2,
legend style={at={(0.02,0.85)},anchor=west}
]
\addplot[color=red,mark=x,solid] table[x index=0, y index=1] {Data_Simu_4/MSEa_dat/CT_MSEa-2/MP.dat};
\addplot[color=blue,mark=x,solid] table[x index=0, y index=1] {Data_Simu_4/MSEa_dat/CT_MSEa-2/BMP.dat};
\addplot[color=red,mark=+,solid] table[x index=0, y index=1] {Data_Simu_4/MSEa_dat/CT_MSEa-2/OMP.dat};
\addplot[color=blue,mark=+,solid] table[x index=0, y index=1] {Data_Simu_4/MSEa_dat/CT_MSEa-2/BOMP.dat};
\addplot[color=red,mark=o,solid] table[x index=0, y index=1] {Data_Simu_4/MSEa_dat/CT_MSEa-2/StOMP.dat};
\addplot[color=blue,mark=o,solid] table[x index=0, y index=1] {Data_Simu_4/MSEa_dat/CT_MSEa-2/BStOMP.dat};
\addplot[color=red,mark=triangle,solid] table[x index=0, y index=1] {Data_Simu_4/MSEa_dat/CT_MSEa-2/SP.dat};
\addplot[color=blue,mark=triangle,solid] table[x index=0, y index=1] {Data_Simu_4/MSEa_dat/CT_MSEa-2/BSP.dat};
\addplot[color=cyan,mark=star,solid] table[x index=0, y index=1] {Data_Simu_4/MSEa_dat/CT_MSEa-2/BPD.dat};
\addplot[color=green,mark=star,solid] table[x index=0, y index=1] {Data_Simu_4/MSEa_dat/CT_MSEa-2/SOBAP.dat};


\legend{MP,BMP,OMP,BOMP,StOMP,BStOMP,SP,BSP,BPD,SOBAP}
\end{axis}
\end{tikzpicture}

\vspace{0.3cm}
\begin{tikzpicture}
\begin{axis}[
xmin=0,
xmax=1,
ymin=0,
ymax=1,
xlabel=$N/M$,
ylabel=$K/N$, 
grid=major,
legend columns=2,
legend style={at={(0.02,0.85)},anchor=west}
]
\addplot[color=red,mark=x,solid] table[x index=0, y index=1] {Data_Simu_4/Pe_cw_dat/CT_Pe_cw-2/MP.dat};
\addplot[color=blue,mark=x,solid] table[x index=0, y index=1] {Data_Simu_4/Pe_cw_dat/CT_Pe_cw-2/BMP.dat};
\addplot[color=red,mark=+,solid] table[x index=0, y index=1] {Data_Simu_4/Pe_cw_dat/CT_Pe_cw-2/OMP.dat};
\addplot[color=blue,mark=+,solid] table[x index=0, y index=1] {Data_Simu_4/Pe_cw_dat/CT_Pe_cw-2/BOMP.dat};
\addplot[color=red,mark=o,solid] table[x index=0, y index=1] {Data_Simu_4/Pe_cw_dat/CT_Pe_cw-2/StOMP.dat};
\addplot[color=blue,mark=o,solid] table[x index=0, y index=1] {Data_Simu_4/Pe_cw_dat/CT_Pe_cw-2/BStOMP.dat};
\addplot[color=red,mark=triangle,solid] table[x index=0, y index=1] {Data_Simu_4/Pe_cw_dat/CT_Pe_cw-2/SP.dat};
\addplot[color=blue,mark=triangle,solid] table[x index=0, y index=1] {Data_Simu_4/Pe_cw_dat/CT_Pe_cw-2/BSP.dat};

\addplot[color=green,mark=star,solid] table[x index=0, y index=1] {Data_Simu_4/Pe_cw_dat/CT_Pe_cw-2/SOBAP.dat};


\legend{MP,BMP,OMP,BOMP,StOMP,BStOMP,SP,BSP,SOBAP}

\end{axis}
\end{tikzpicture}
 \caption{Phase transition curves for MSE=$10^{-2}$ (top) and $P_e=10^{-2}$ (bottom).}\label{fig:PTuniform}
\end{figure}

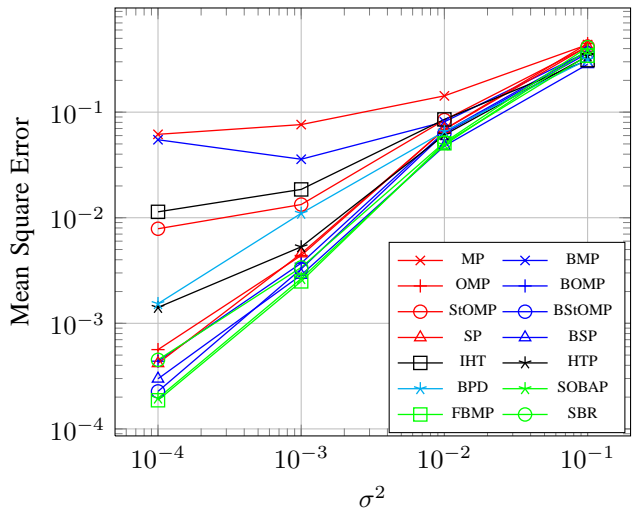
\begin{figure} 

\begin{tikzpicture}
\pgfplotstableread{Data_Simu_6/FMvsSNR_dims256_vara1_gaussian_random_3000/rho0.6_delta0.35/MSEa.dat}\loadedtable
\begin{loglogaxis}[
xlabel=$\sigma^2$,
ylabel=Mean Square Error, 
grid=major,
legend columns=2,
legend style={at={(0.99,0.23)},anchor=east}
]
\addplot[color=red,mark=x,solid] table[x index=0, y index=1] {\loadedtable};
\addplot[color=blue,mark=x] table[x index=0, y index=5] {\loadedtable};
\addplot[color=red,mark=+,solid] table[x index=0, y index=2] {\loadedtable};
\addplot[color=blue,mark=+] table[x index=0, y index=6] {\loadedtable};
\addplot[color=red,mark=o,solid] table[x index=0, y index=3] {\loadedtable};
\addplot[color=blue,mark=o] table[x index=0, y index=7] {\loadedtable};
\addplot[color=red,mark=triangle,solid] table[x index=0, y index=4] {\loadedtable};
\addplot[color=blue,mark=triangle] table[x index=0, y index=8] {\loadedtable};
\addplot[color=black,mark=square] table[x index=0, y index=9] {\loadedtable};
\addplot[color=black,mark=star] table[x index=0, y index=10] {\loadedtable};
\addplot[color=cyan,mark=star] table[x index=0, y index=11] {\loadedtable};
\addplot[color=green,mark=star] table[x index=0, y index=12] {\loadedtable};
\addplot[color=green,mark=square] table[x index=0, y index=13] {\loadedtable};
\addplot[color=green,mark=o] table[x index=0, y index=14] {\loadedtable};
\legend{MP,BMP,OMP,BOMP,StOMP,BStOMP,SP,BSP,IHT,HTP,BPD,SOBAP,FBMP,SBR}

\end{loglogaxis}
\end{tikzpicture}
 \caption{MSE versus $\sigma^2$ for $K=54$, $N=154$ and $M=256$.}\label{fig:MSEvsNoise}
\end{figure}


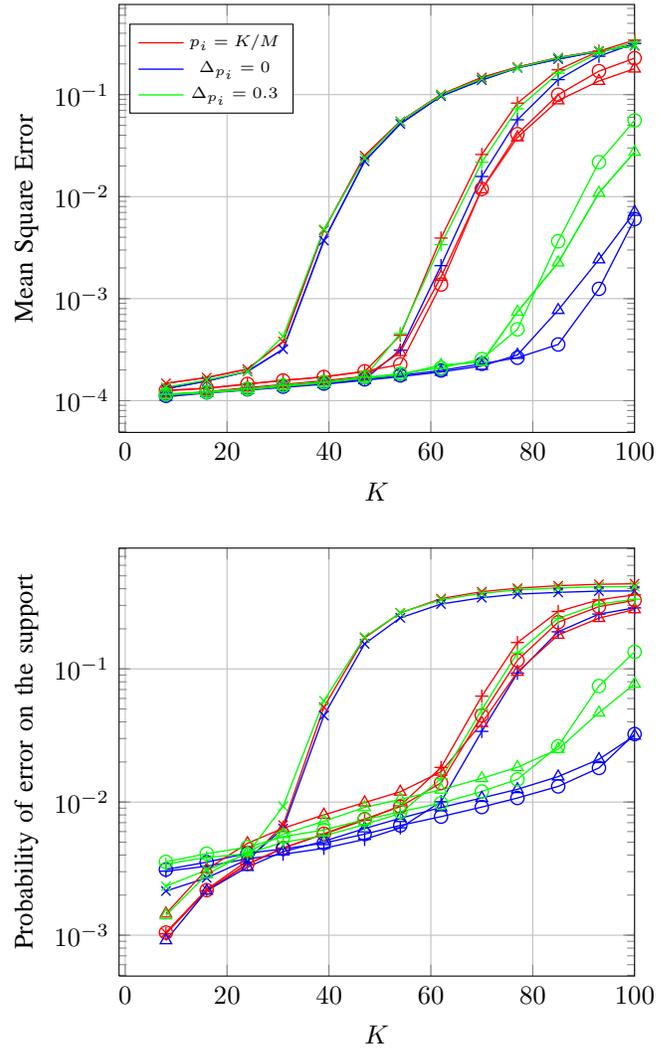
\begin{figure} 

\begin{tikzpicture}
\begin{semilogyaxis}[
xmax=100,
xlabel=$K$,
ylabel=Mean Square Error, 
grid=major,
legend columns=1,
legend style={at={(0.02,0.85)},anchor=west}
]
\addplot[color=red,solid] table[x index=0, y index=5] {Data_Simu_4/MSEa_dat/MSEa_rho0.6.dat}; 
\addplot[color=blue] table[x index=0, y index=1] {Data_Simu_5/deltap0/MSEa_dat/MSEa_rho0.6.dat}; 
\addplot[color=green] table[x index=0, y index=4] {Data_Simu_5/deltap0.3/MSEa_dat/MSEa_rho0.6.dat}; 

\addplot[color=red,mark=x,solid] table[x index=0, y index=5] {Data_Simu_4/MSEa_dat/MSEa_rho0.6.dat};
\addplot[color=blue,mark=x] table[x index=0, y index=1] {Data_Simu_5/deltap0/MSEa_dat/MSEa_rho0.6.dat};
\addplot[color=red,mark=+,solid] table[x index=0, y index=6] {Data_Simu_4/MSEa_dat/MSEa_rho0.6.dat};
\addplot[color=blue,mark=+] table[x index=0, y index=2] {Data_Simu_5/deltap0/MSEa_dat/MSEa_rho0.6.dat};
\addplot[color=red,mark=o,solid] table[x index=0, y index=7] {Data_Simu_4/MSEa_dat/MSEa_rho0.6.dat};
\addplot[color=blue,mark=o] table[x index=0, y index=3] {Data_Simu_5/deltap0/MSEa_dat/MSEa_rho0.6.dat};
\addplot[color=red,mark=triangle,solid] table[x index=0, y index=8] {Data_Simu_4/MSEa_dat/MSEa_rho0.6.dat};
\addplot[color=blue,mark=triangle] table[x index=0, y index=4] {Data_Simu_5/deltap0/MSEa_dat/MSEa_rho0.6.dat};

\addplot[color=green,mark=x] table[x index=0, y index=1] {Data_Simu_5/deltap0.3/MSEa_dat/MSEa_rho0.6.dat};
\addplot[color=green,mark=+] table[x index=0, y index=2] {Data_Simu_5/deltap0.3/MSEa_dat/MSEa_rho0.6.dat};
\addplot[color=green,mark=o] table[x index=0, y index=3] {Data_Simu_5/deltap0.3/MSEa_dat/MSEa_rho0.6.dat};
\addplot[color=green,mark=triangle] table[x index=0, y index=4] {Data_Simu_5/deltap0.3/MSEa_dat/MSEa_rho0.6.dat};


\legend{$p_i=K/M$,$\Delta_{p_i}=0$, $\Delta_{p_i}=0.3$}

\end{semilogyaxis}
\end{tikzpicture}

\vspace{0.5cm}

\begin{tikzpicture}
\begin{semilogyaxis}[
xmax=100,
xlabel=$K$,
ylabel=Probability of error on the support, 
grid=major,
legend columns=2,
legend style={at={(0.02,0.83)},anchor=west}]
\addplot[color=red,mark=x,solid] table[x index=0, y index=5] {Data_Simu_4/Pe_cw_dat/Pe_cw_rho_0.6.dat};
\addplot[color=blue,mark=x] table[x index=0, y index=1] {Data_Simu_5/deltap0/Pe_cw_dat/Pe_cw_rho_0.6.dat};
\addplot[color=red,mark=+,solid] table[x index=0, y index=6] {Data_Simu_4/Pe_cw_dat/Pe_cw_rho_0.6.dat};
\addplot[color=blue,mark=+] table[x index=0, y index=2] {Data_Simu_5/deltap0/Pe_cw_dat/Pe_cw_rho_0.6.dat};
\addplot[color=red,mark=o,solid] table[x index=0, y index=7] {Data_Simu_4/Pe_cw_dat/Pe_cw_rho_0.6.dat};
\addplot[color=blue,mark=o] table[x index=0, y index=3] {Data_Simu_5/deltap0/Pe_cw_dat/Pe_cw_rho_0.6.dat};
\addplot[color=red,mark=triangle,solid] table[x index=0, y index=8] {Data_Simu_4/Pe_cw_dat/Pe_cw_rho_0.6.dat};
\addplot[color=blue,mark=triangle] table[x index=0, y index=4] {Data_Simu_5/deltap0/Pe_cw_dat/Pe_cw_rho_0.6.dat};

\addplot[color=green,mark=x] table[x index=0, y index=1] {Data_Simu_5/deltap0.3/Pe_cw_dat/Pe_cw_rho_0.6.dat};

\addplot[color=green,mark=+] table[x index=0, y index=2] {Data_Simu_5/deltap0.3/Pe_cw_dat/Pe_cw_rho_0.6.dat};

\addplot[color=green,mark=o] table[x index=0, y index=3] {Data_Simu_5/deltap0.3/Pe_cw_dat/Pe_cw_rho_0.6.dat};

\addplot[color=green,mark=triangle] table[x index=0, y index=4] {Data_Simu_5/deltap0.3/Pe_cw_dat/Pe_cw_rho_0.6.dat};


\end{semilogyaxis}
\end{tikzpicture}

%
%

 \caption{MSE and probability of error on the support versus the number of non-zero coefficients $K$ for BMP ($\times$), BOMP (+), BStOMP ($\circ$) and BSP ($\triangle$).}
 \label{fig:NonUniformvsK}
\end{figure}

\begin{figure} 
\begin{tikzpicture}
\begin{axis}[
xmin=0,
xmax=1,
ymin=0,
ymax=1,
xlabel=$N/M$,
ylabel=$K/N$, 
grid=major,
legend columns=1,
legend style={at={(0.02,0.85)},anchor=west}
]
\addplot[color=red] table[x index=0, y index=1] {Data_Simu_4/MSEa_dat/CT_MSEa-2/BMP.dat}; 
\addplot[color=blue] table[x index=0, y index=1] {Data_Simu_5/deltap0/MSEa_dat/CT_MSEa-2/BMP.dat}; 

\addplot[color=red,mark=x,solid] table[x index=0, y index=1] {Data_Simu_4/MSEa_dat/CT_MSEa-2/BMP.dat};
\addplot[color=blue,mark=x,solid] table[x index=0, y index=1] {Data_Simu_5/deltap0/MSEa_dat/CT_MSEa-2/BMP.dat};
\addplot[color=red,mark=+,solid] table[x index=0, y index=1] {Data_Simu_4/MSEa_dat/CT_MSEa-2/BOMP.dat};
\addplot[color=blue,mark=+,solid] table[x index=0, y index=1] {Data_Simu_5/deltap0/MSEa_dat/CT_MSEa-2/BOMP.dat};
\addplot[color=red,mark=o,solid] table[x index=0, y index=1] {Data_Simu_4/MSEa_dat/CT_MSEa-2/BStOMP.dat};
\addplot[color=blue,mark=o,solid] table[x index=0, y index=1] {Data_Simu_5/deltap0/MSEa_dat/CT_MSEa-2/BStOMP.dat};
\addplot[color=red,mark=triangle,solid] table[x index=0, y index=1] {Data_Simu_4/MSEa_dat/CT_MSEa-2/BSP.dat};
\addplot[color=blue,mark=triangle,solid] table[x index=0, y index=1] {Data_Simu_5/deltap0/MSEa_dat/CT_MSEa-2/BSP.dat};


\legend{$p_i=K/M$,$\Delta_{p_i}=0$}
\end{axis}
\end{tikzpicture}

\begin{tikzpicture}
\begin{axis}[
xmin=0,
xmax=1,
ymin=0,
ymax=1,
xlabel=$N/M$,
ylabel=$K/N$, 
grid=major,
legend columns=2,
legend style={at={(0.02,0.85)},anchor=west}
]
\addplot[color=red,mark=x,solid] table[x index=0, y index=1] {Data_Simu_4/Pe_cw_dat/CT_Pe_cw-2/BMP.dat};
\addplot[color=blue,mark=x,solid] table[x index=0, y index=1] {Data_Simu_5/deltap0/Pe_cw_dat/CT_Pe_cw-2/BMP.dat};
\addplot[color=red,mark=+,solid] table[x index=0, y index=1] {Data_Simu_4/Pe_cw_dat/CT_Pe_cw-2/BOMP.dat};
\addplot[color=blue,mark=+,solid] table[x index=0, y index=1] {Data_Simu_5/deltap0/Pe_cw_dat/CT_Pe_cw-2/BOMP.dat};
\addplot[color=red,mark=o,solid] table[x index=0, y index=1] {Data_Simu_4/Pe_cw_dat/CT_Pe_cw-2/BStOMP.dat};
\addplot[color=blue,mark=o,solid] table[x index=0, y index=1] {Data_Simu_5/deltap0/Pe_cw_dat/CT_Pe_cw-2/BStOMP.dat};
\addplot[color=red,mark=triangle,solid] table[x index=0, y index=1] {Data_Simu_4/Pe_cw_dat/CT_Pe_cw-2/BSP.dat};
\addplot[color=blue,mark=triangle,solid] table[x index=0, y index=1] {Data_Simu_5/deltap0/Pe_cw_dat/CT_Pe_cw-2/BSP.dat};



\end{axis}
\end{tikzpicture}

 \caption{Phase transition curves for MSE=$10^{-2}$ (top) and $P_e=10^{-2}$ (bottom) for BMP ($\times$), BOMP (+), BStOMP ($\circ$) and BSP ($\triangle$).} \label{fig:PTnonuniform}
\end{figure}
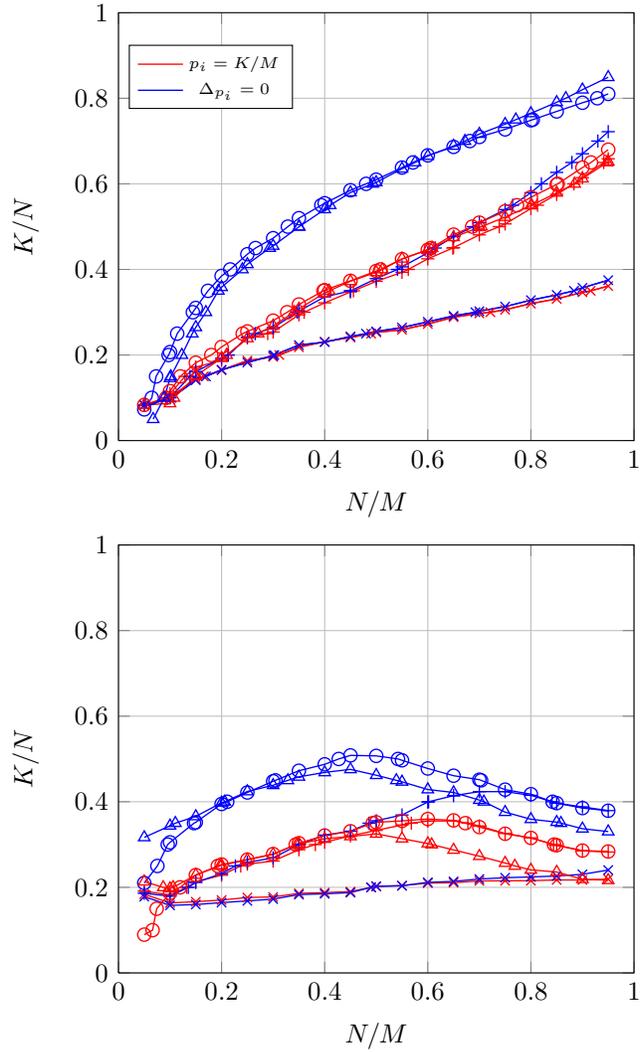

\end{document}